\documentclass[twocolumn,aps,showpacs,prb,tightenlines,amsmath,amssymb,superscriptaddress,10pt, floatfix]{revtex4-1}
\usepackage{graphicx}
\usepackage{amssymb}
\usepackage{dcolumn}
\usepackage{amsmath}
\usepackage{bm}
\usepackage{colordvi}
\usepackage{float}
\usepackage{color}
\usepackage{hyperref}
\usepackage[section]{placeins}

\begin{document}

\title{Quasiparticle states in AC-driven quantum conductors: A scattering theory approach}
\author{Y. Yin}
\thanks{Author to whom correspondence should be addressed}
\email{yin80@scu.edu.cn.}
\affiliation{Laboratory of Mesoscopic and Low Dimensional Physics,
Department of Physics, Sichuan University, Chengdu, Sichuan, 610064, China}
\date{\today}

\begin{abstract}
We introduce a general procedure to extract the wave function of quasiparticles in AC-driven quantum conductors. By incorporating the Bloch-Messiah reduction into the scattering theory approach to quantum transport, we construct the many-body state from the scattering matrix, within which the wave function of quasiparticles can be extracted. We find that two species of quasiparticles can be excited, while both of them are superpositions of particle and hole states in the Fermi sea. Due to the electron number conservation, the quasiparticles are always created in pairs, resulting in particle-hole pair excitations in quantum conductors. We apply our approach to compare these particle-hole pairs excited in perfectly transparent quantum contacts and mesoscopic capacitors with single-level quantum dot. For the quantum contacts, particle-hole pairs with multiple modes can be excited, while the excitation probability of each mode increases as the strength of the driving field increasing. The particle(hole) components are always in phase (180 degrees out of phase) with the driving field. These features agree with the results obtained via the extended Keldysh-Green's function technique in previous works.  For the mesoscopic capacitors, only one mode of particle-hole pairs can be excited, while the corresponding excitation probability undergoes an oscillation as the strength of the driving field increasing. A nonzero phase delay between the particle(hole) component and the driving field exists, which can be tuned via the dot-lead coupling in the mesoscopic capacitor. These features are very distinct from the case of quantum contacts, which can be attributed to the finite dwell time of electrons in the mesoscopic capacitor. We also find that our approach can also offer a general way to extract the wave function of quasiparticles from the first-order electronic correlation function, making it helpful for the signal processing of relevant experiments in electron quantum optics.
\end{abstract}


\maketitle

\section{Introduction}
\label{sec1}

In recent decades, there has been growing interest in the emerging field of electron quantum optics, which focus on the
understanding the behavior of electrons in ballistic quantum conductors following the paradigms of quantum
optics.\cite{grenier2011, bocquillon2014} As has been pointed out by scattering theory approach of quantum transport,
quantum electric currents can be regarded as been built from elementary excitations in the Fermi sea of quantum
conductors.\cite{landauer1985, buttiker1985, buttiker1986, martin1992, lesovik2011} Previous research have focus on the
statistic properties of these elementary excitations, which can be extracted from the correlation functions of the
current in the DC case.\cite{levitov1996, blanter2001} Electronic Mach-Zehnder\cite{Ji2003, neder2007, roulleau2007,
  roulleau2008} and Hanbury-Brown-Twiss\cite{liu1998, henny1999, oliver1999} interferometry have been realized in the
quantum Hall edge channels, revealing the long coherence length, antibunching and entanglement of elementary excitations
in such system.\cite{beenakker2003, samuelsson2004, haack2011, sherkunov2012, hofer2017}

Going beyond the DC case, recent developments in on-demand electron sources\cite{feve2007, mahe2010, dubois2013,
  fletcher2013} and quantum tomographic methods\cite{grenier2011a, bocquillon2013, jullien2014} show that individual
excitations in quantum conductors can be generated and detected under properly engineered time-dependent driving, paving
the way for electron quantum optics to the single electron level.\cite{roussel2017, marguerite2017} Hence, besides the
statistic description widely used in the DC case, an alternative description suitable for such time-dependent case is
preferred. In particular, it is natural to ask what is the wave function of these excitations, which is one of the most straightforward way to characterize them.

Generally speaking, the elementary excitation in the Fermi sea are quasiparticles composed of electrons and/or
holes.\cite{rychkov2005} The time-dependent driving mixes them with different energies, makes it a non-trivial task to
find the wave function of these quasiparticles even in the absence of interaction.\cite{vanevi2017} For quantum contacts
with weak harmonic driving, the wave function has been addressed perturbatively, which is valid in the case when the
amplitude of the driving field is much smaller than the driving frequency.\cite{moskalets2002, rychkov2005,
  polianski2005} In the case of strong harmonic driving, the wave function can be obtained by examining the cumulant
generating function of the transferred charge in quantum contacts, which has been obtained by using the extended
Keldysh-Green's function technique.\cite{vanevi2007, vanevi2008, vanevi2012, vanevi2016}. For specific case when only
electrons(or holes) are excited in the Fermi sea, the wave function of the quasiparticles, now called levitons, can
also be constructed beyond the perturbation theory.\cite{keeling2006, keeling2008, sherkunov2009, zhang2009,
  dasenbrook2015}

Despite these studies which have improved our understanding, a general and explicit relation between the wave function
of the quasiparticles and the scattering matrix, which plays a central role in the scattering theory approach, is still
missing. Due to the success of the scattering theory approach, such relation is expected to offer an efficient and transparent way to study the properties of these quasiparticles. In fact, the analogy between the quantum transport of electrons and the propagation of photons has already been emphasized by the scattering theory approach in the early stage.\cite{martin1992, buttiker1992, imry1999} Recently, first-order electronic correlation functions,\cite{haack2013, moskalets2015, moskalets2017} or equivalently, electronic Wigner functions,\cite{grenier2011a, ferraro2013, jullien2014, kashcheyevs2017} have been introduced within such approach, from which the information of these quasiparticles can be extracted. These studies have paved the way along this direction.

Hence the goal of this paper is to fill this gap and provide a general approach to obtain the wave function of the quasiparticles in AC-driven quantum conductors, which is widely studied in the field of electron quantum optics. We will concentrate on the case without interaction, which is valid in the integer quantum Hall regime where most of the electron quantum optics experiments are performed.\cite{bocquillon2014} We find that, by incorporating the Bloch-Messiah reduction\cite{bloch1962} into the scattering theory approach, the many-body state of the electrons for quantum conductors in the zero-temperature limit can be constructed as
\begin{eqnarray}
  | \Psi_b \rangle & = & \prod_{\rm \delta \in [0, 1)} 
                         \prod_{\rm k = 1,2,...} \frac{\gamma^{\dagger}_{\rm k -}(\delta) \gamma^{\dagger}_{\rm k +}(\delta)}{i \sqrt{p_k(\delta)}}
                          | F \rangle,
                         \label{s1:eq1}
\end{eqnarray}
with $| F \rangle$ representing the unperturbed Ferm sea. Such many-body wave function suggests that two species of quasiparticles($+/-$), which represented by the creation operators $\gamma^{\dagger}_{\rm k s}(\delta)$($s=+/-$), are excited during the scattering. The corresponding creation operators $\gamma^{\dagger}_{\rm k s}(\delta)$ have the form
\begin{eqnarray}
  \gamma^{\dagger}_{\rm k -}(\delta) & = & - \sqrt{1 - p_k(\delta)} B_{\rm p k}(\delta) +
                                   i \sqrt{p_k(\delta)} B^{\dagger}_{\rm h k}(\delta),
                                   \nonumber\\
  \gamma^{\dagger}_{\rm k +}(\delta) & = & - \sqrt{1 - p_k(\delta)} B_{\rm h k}(\delta) -
                                   i \sqrt{p_k(\delta)} B^{\dagger}_{\rm p k}(\delta),
  \label{s1:eq2}
\end{eqnarray}
with $B_{\rm p(h) k}(\delta)$ representing the particle(hole) components of the wave function for the quasiparticles. 

The above expression indicates that the quasiparticle, which represented by the creation operator $\gamma^{\dagger}_{\rm k s}(\delta)$($s=+/-$), increases/decreases the total electron number by one. Due to the electron number conservation, they have to be created in pairs, leading to particle-hole pair excitations in the quantum conductor. The corresponding excitation probability $p_k(\delta)$ do not depend on the index of the species $s$ and is only a function of the mode index $k$ and the parameter $\delta$. Both the particle(hole) components and the excitation probability can be decided from the singular value decomposition (SVD) for sub-matrices of the scattering matrix, which offers a general way to extract the quasiparticle wave function from the scattering matrix of the quantum conductors.

To better demonstrate our approach, two specific cases are studied in this paper: (1) a perfectly transparent quantum contact and (2) a mesoscopic capacitor. The case of the perfectly transparent quantum contact has been addressed via the extended Keldysh-Green's function technique in previous works.\cite{vanevi2016, vanevi2017} In this case, particle-hole pairs with different modes($k=1,2,...$) can be created during the scattering and the excitation probability of each mode increases with the strength of the driving field increasing. While the particle components of the wave function stay in phase with the driving field, the hole components are always 180 degrees out of phase. We find that both the excitation probability and the particle(hole) components of the wave function obtained from our approach agrees quiet well with the ones reported in previous works via the extended Keldysh-Green's function technique. 

We then turn to study the wave function of quasiparticles created in the mesoscopic capacitor, which has not been explicitly given in a general case in previous studies. We find that in this case, only one mode of particle-hole pairs can be excited, whose excitation probability undergoes an oscillation as the strength of the driving field increasing. Moreover, a phase delay between the particle(hole) components of the wave function and the driving field exists. These two features are very distinct from the case of the quantum contact, which can be attributed to the finite dwell time of electrons in the mesoscopic capacitor.

Although our approach mainly concentrates on the theoretically aspect, it can also be helpful for the signal processing techniques in experiments. We find that the our approach offers a systematic way to extract the wave function of quasiparticles from the first-order electronic correlation function, or equivalently, from the electronic Wigner function, which generalizes the signal processing techniques introduced by Marguerite {\em et al.} in Ref.~[\onlinecite{marguerite2017}]. 

The paper is organized as follows. In Sec.~\ref{sec2}, we present our approach by using the two-terminal quantum conductor as an example. Then, we apply our approach to study the case of the perfectly transparent quantum contact and the mesoscopic  capacitor in Sec.~\ref{sec3} and~\ref{sec4}, respectively. The procedure to extract the wave function of the quasiparticles from the first-order electronic correlation function is discussed in Sec.~\ref{sec5}. We summarize our conclusions in Sec.~\ref{sec6}.

\section{Wave function of elementary excitations from scattering matrix}
\label{sec2}

In this section, we demonstrate our approach in a typical two-terminal AC-driven quantum conductor. We show that the elementary excitations are particle-hole pairs composed of two species of quasiparticles, whose wave function can be solely decided by the scattering matrix of the quantum conductor.

\begin{figure}
  \centering
  \includegraphics[width=7.0cm]{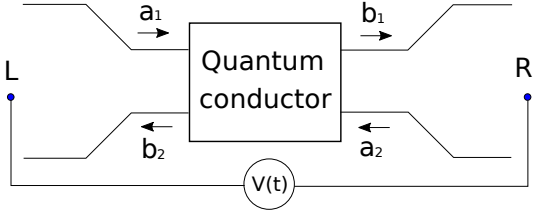}
  \caption{ Schematic of the two-terminal quantum conductor driven by the time-dependent bias voltage $V(t)$.}
  \label{fig1}
\end{figure}

We start our discuss by considering the two-terminal quantum conductor as illustrated in Fig.~\ref{fig1}, which
connects two reservoir L and R via two ideal leads. For simplicity, we assume each lead contain two channels $1$ and
$2$, sustaining the left- and right-moving electron states, respectively. Both reservoirs are assumed to be in local
thermal equilibrium with the same temperature $T$ and Fermi energy $E_F$, A time-dependent bias voltage $V(t)$ is applied between the reservoir, leading to quantum electric currents in the quantum conductor. Following previous works,\cite{vanevi2016} we concentrate on the zero-temperature limit and non-interacting case in the following discussion.

By introducing creation[annihilation] operators of electrons for incoming $a^{\dagger}_1(E)$, $a^{\dagger}_2(E)$[$a_1(E)$,
$a_2(E)$] and outgoing $b^{\dagger}_1(E)$, $b^{\dagger}_2(E)$[$b_1(E)$, $b_2(E)$] channels, the scattering
matrix in the energy domain can be given as
\begin{eqnarray}
  b_{\lambda}(E) = \sum_{\rm \eta=1,2} \int \frac{dE'}{2\pi} S_{\rm \lambda \eta}(E, E') a_{\eta}(E'),
  \label{s2:eq1}
\end{eqnarray}
where the off-diagonal element of the scattering matrix[$S_{\rm \lambda \eta}(E, E')$, with $E \ne E'$] is due to the
time-dependent driving $V(t)$. For typical driving with fundamental frequency $\Omega_0$, it
is traditionally to work in the Floquet space, where Eq.~\eqref{s2:eq1} has the form
\begin{eqnarray}
  b_{\rm \lambda n}(\delta) = \sum_{\rm \eta=1,2} \sum_m S_{\rm \lambda \eta, n m}(\delta) a_{\rm \eta m}(\delta),
  \label{s2:eq2}
\end{eqnarray}
with the Floquet scattering matrix defined as
\begin{eqnarray}
  S_{\rm \lambda \eta, n m}(\delta) & = & \frac{2\pi}{\Omega_0} S_{\rm \lambda \eta}(n\Omega_0 + \delta\Omega_0, m\Omega_0 + \delta\Omega_0).
  \label{s2:eq2-1}
\end{eqnarray}
Note that we have defined the electron annihilation operator in the Floquet space as $a_{\rm \lambda n}(\delta) = a_{\lambda}(\delta\Omega_0 + n\Omega_0)$[$b_{\rm \lambda n}(\delta) = b_{\lambda}(\delta\Omega_0 + n \Omega_0)$] with $\lambda=1, 2$, $n \in \mathbb{Z}$ and $\delta \in [0, 1)$. Also note that we have chosen the natural units throughout this paper.

For the problem we study here, it is convenient to present the Floquet scattering matrix in a particle-hole basis. To do
so, we introduce a canonical particle-hole transformation
\begin{eqnarray}
  b_{\rm \lambda n}(\delta) & = & \left\{
                                    \begin{tabular}{cc}
                                      $b^p_{\rm \lambda n}(\delta)$, & $n \ge 0$\\
                                      $[b^h_{\lambda n}(\delta)]^{\dagger}$, &
                                      $n<0$\\
                                    \end{tabular}
  \right.
  \label{s2:eq3}
\end{eqnarray}
for the outgoing electron along the channel $\lambda$ ($\lambda=1, 2$), where $b^p_{\rm \lambda n}(\delta)$ and
$b^h_{\rm \lambda n}(\delta)$ represent the annihilation operator for particle and hole states, respectively. The
same transformation can be applied to the incoming electrons in the same way, leading to
$a^p_{\rm \lambda n}(\delta)$[$a^h_{\rm \lambda n}(\delta)$] representing the particle[hole] state for the
corresponding incoming state.

The Floquet scattering matrix in such particle-hole basis can be decomposed as
\begin{eqnarray}
  \left[\begin{tabular}{c}
          $\vec{b}_p(\delta)$\\
          $\vec{b}^{\ast}_h(\delta)$
        \end{tabular}\right] & = & \left[\begin{tabular}{cc}
                                           $S^{\rm pp}(\delta)$ & $S^{\rm ph}(\delta)$\\
                                           $S^{\rm hp}(\delta)$ & $S^{\rm hh}(\delta)$\\
                                         \end{tabular}\right] \left[\begin{tabular}{c}
                                                                      $\vec{a}_p(\delta)$\\
                                                                      $\vec{a}^{\ast}_h(\delta)$
                                                                    \end{tabular}\right],
  \label{s2:eq4}
\end{eqnarray}
where the vectors of operators are defined as
\begin{eqnarray}
  \vec{b}_p(\delta) & = & \left[\begin{tabular}{c}
                                    $\vdots$\\
                                    $b^p_{\rm 1 n}(\delta)$\\
                                    $\vdots$\\                                    
                                    $b^p_{\rm 2 n}(\delta)$\\                                    
                                    $\vdots$
                                  \end{tabular}\right], n \ge 0,\nonumber\\
  \vec{b}^{\ast}_h(\delta) & = & \left[\begin{tabular}{c}
                                           $\vdots$\\                                
                                           $[b^h_{\rm 1 n}(\delta)]^{\dagger}$\\
                                           $\vdots$\\                                           
                                           $[b^h_{\rm 2 n}(\delta)]^{\dagger}$\\
                                           $\vdots$                                
                                         \end{tabular}\right], m < 0,
  \label{s2:eq5}
\end{eqnarray}
while $\vec{a}_p(\delta)$ and $\vec{a}^{\ast}_h(\delta)$ can be defined in the same way. Note that the four
sub-matrices in Eq.~\eqref{s2:eq4} can be related to the Floquet scattering matrix as
\begin{eqnarray}
  S^{\rm pp}_{\rm \lambda\eta, nn'}(\delta) & = & S_{\rm \lambda\eta, nn'}(\delta), n, n' \ge 0\nonumber\\
  S^{\rm ph}_{\rm \lambda\eta, nn'}(\delta) & = & S_{\rm \lambda\eta, nn'}(\delta), n \ge 0, n' < 0\nonumber\\
  S^{\rm hp}_{\rm \lambda\eta, nn'}(\delta) & = & S_{\rm \lambda\eta, nn'}(\delta), n < 0, n' \ge 0\nonumber\\
  S^{\rm hh}_{\rm \lambda\eta, nn'}(\delta) & = & S_{\rm \lambda\eta, nn'}(\delta), n, n' < 0.
  \label{s2:eq5-1}
\end{eqnarray}

It is worth noting that since the scattering matrix is unitary, it can be proved that (see Appendix~\ref{app1} for details), the SVD of the four sub-matrices in Eq.~\eqref{s2:eq4} satisfy the following relation
\begin{eqnarray}
  S^{\rm pp}(\delta) & = & U^p(\delta) P(\delta) [V^p(\delta)]^{\dagger}, \nonumber\\
  S^{\rm hh}(\delta) & = & U^h(\delta) P(\delta) [V^h(\delta)]^{\dagger}, \nonumber\\
  S^{\rm ph}(\delta) & = & U^p(\delta) i \sqrt{ I - P^2(\delta) } [V^h(\delta)]^{\dagger}, \nonumber\\
  S^{\rm hp}(\delta) & = & U^h(\delta) i \sqrt{ I - P^2(\delta) } [V^p(\delta)]^{\dagger},
  \label{s2:eq5-2}
\end{eqnarray}
where $I$ is a unit matrix. $P(\delta)$ is the diagonal matrix whose diagonal entry $p_k(\delta)$ is the $k$th
singular value of the matrix $S^{\rm pp}(\delta)$[$S^{\rm hh}(\delta)$]. The corresponding left and right singular
vectors are stored in the $k$th-column of the unity matrices $U^{\rm p(h)}$ and $V^{\rm p(h)}$, respectively.

Now we turn to discuss the many-body state, from which the wave function of the quasiparticle can be
extracted. According to the assumption of the scattering theory approach, the many-body wave function of the system can
be written as a direct product of the incoming and outgoing states of the form
\begin{eqnarray}
  | \Psi_{\rm sys} \rangle & = & | \Psi_a \rangle \otimes | \Psi_b \rangle
  \label{s2:eq6}
\end{eqnarray}
in the zero-temperature limit. The incoming state $| \Psi_a \rangle$ characterizes the emission of electrons from the
reservoir L(R) along the channel $1$($2$), which has the form
\begin{eqnarray}
  | \Psi_a \rangle & = & | F_{\rm L} \rangle \otimes | F_{\rm R} \rangle,
  \label{s2:eq7}
\end{eqnarray}
where $| F_{\rm L(R)} \rangle$ represents the Fermi sea corresponding to the reservoir L(R). They can be constructed from the corresponding vacuum state of electrons $| 0_{\rm L(R)} \rangle$ as
\begin{eqnarray}
  | F_{\rm L(R)} \rangle & = & \int^1_0 d\delta \sum_{m<0}
                               a^{\dagger}_{\rm 1(2) m}(\delta) | 0_{\rm L(R)}
                               \rangle,
                               \label{s2:eq8}
\end{eqnarray}
in the Floquet basis. Note that we have chosen the zero of energy so that the Fermi energy $E_F=0$.

The outgoing state $| \Psi_b \rangle$ contains the information of the quasiparticles created during the scattering,
while the Fermi sea states $| F_{\rm L} \rangle \otimes | F_{\rm R} \rangle$ serves as the vacuum state of these
quasiparticles. Although one can construct $| \Psi_b \rangle$ from the vacuum state of electrons
$| 0_{\rm L(R)} \rangle$ by using the scattering matrix given in Eq.~\eqref{s2:eq2}, it is general difficult to obtain
the wave function of the quasiparticles from such expression, except for some specific case, such as weak harmonic
driving and/or leviton excitation.\cite{moskalets2002, rychkov2005, polianski2005, keeling2006, keeling2008,
  sherkunov2009, zhang2009, dasenbrook2015} Alternatively, it is more convenient to extract the information of the
quasiparicles from the pair correlation function,\cite{beenakker2005} since for the non-interacting case considered
here, the many-body state remains Gaussian, whose high order correlation function can be fully constructed from the pair
ones.\cite{cheong2004, corney2006, corney2006a}

Following this idea, we construct the outgoing state $| \Psi_b \rangle$ from the pair correlation function in the
particle-hole basis by using the Bloch-Messiah reduction,\cite{bloch1962} which has been used in the field of quantum information.\cite{kraus2009, weedbrook2012} To make our discussion transparent, we only present the main result here, leaving the details of the derivation in Appendix~\ref{app2}.

We find that the outgoing state $| \Psi_b \rangle$ can be written as
\begin{eqnarray}
  \hspace{-0.5cm}| \Psi_b \rangle & = & \prod_{\rm \delta \in [0, 1)}
                         \prod_{k = 1, 2, ...} \Big[ \sqrt{ 1 - p_k(\delta)}
                         \nonumber\\
                   && \hspace{0.5cm}\mbox{}+  i \sqrt{p_k(\delta)}
                      B^{\dagger}_{\rm p k}(\delta)
                      B^{\dagger}_{\rm h k}(\delta) \Big] | F_{\rm L} \rangle
                      \otimes | F_{\rm R} \rangle,
                      \label{s2:eq9}
\end{eqnarray}
where operator $B^{\dagger}_{\rm p(h) k}(\delta)$ represents the creation operator of
the particle(hole) component of the wave function, which has the form
\begin{eqnarray}
  B_{\rm p k}(\delta) & = & \sum_{\rm \lambda=1, 2} \sum_{\rm n \ge 0} [ U^p_{\rm \lambda
                              n, k}(\delta) ]^{\ast}
                              b^p_{\rm \lambda n}(\delta), \nonumber\\
  B_{\rm h k}(\delta) & = & \sum_{\rm \lambda=1, 2} \sum_{\rm n < 0} U^h_{\rm \lambda n,
                              k}(\delta)
                              b^h_{\rm \lambda n}(\delta),
                              \label{s2:eq10}
\end{eqnarray}
with $U^p_{\rm k, \lambda n}(\delta)$[$U^h_{\rm k, \lambda n}(\delta)$] representing the matrix element of the
$k$th column of the left-singular vectors matrix as given in Eq.~\eqref{s2:eq5-2}, while the excitation probability $p_k(\delta)$ is just the corresponding $k$th singular value. Hence, the many-body state can be fully determined by the scattering matrix.

Note that the state given in Eq.~\eqref{s2:eq9} has a similar form as the well-known BCS wave function, so it has been
referred to as Gaussian BCS state.\cite{kraus2009} In fact, the wave function of the quasiparticle can be better
illustrated by performing the Bogoliubov transformation. By introducing $u_k(\delta) = - \sqrt{1 - p_k(\delta)}$ and
$v_k(\delta) = i \sqrt{p_k(\delta)}$, one has
\begin{eqnarray}
  | \Psi_b \rangle & = & \prod_{\rm \delta \in [0, 1)}
                         \prod_{\rm k = 1,2, ...} \frac{\gamma^{\dagger}_{\rm k -}(\delta) \gamma^{\dagger}_{\rm k +}(\delta)}{v_k(\delta)}
                          | F_{\rm L} \rangle \otimes | F_{\rm R} \rangle,
                         \label{s2:eq11}
\end{eqnarray}
where $\gamma^{\dagger}_{\rm k \pm}(\delta)$ are the creation operators representing the quasiparticles, which can be
given as
\begin{eqnarray}
  \gamma^{\dagger}_{\rm k -}(\delta) & = & u_k(\delta) B_{\rm p k}(\delta) +
                                   v_k(\delta) B^{\dagger}_{\rm h k}(\delta),
                                   \nonumber\\
  \gamma^{\dagger}_{\rm k +}(\delta) & = & u_k(\delta) B_{\rm h k}(\delta) -
                                   v_k(\delta) B^{\dagger}_{\rm p k}(\delta).
  \label{s2:eq12}
\end{eqnarray}
Now one can see clearly that during the scattering, two species of quasiparticles are created, which are characterized by the creation operators $\gamma^{\dagger}_{\rm k -}(\delta)$ and $\gamma^{\dagger}_{\rm k +}(\delta)$, respectively. While the former decreases the total electron number by $1$, the latter increases the total electron number by $1$. Due to the electron number conservation, one cannot create a single species of quasiparticle during the scattering. They can only be created in pairs, and hence the corresponding elementary excitations in the AC-driven quantum conductors are just particle-hole pairs composed by these quasiparticles.\footnote{Levitons are a special case, when the particle components of the particle-hole pairs are only created in one channel, while the hole components are only created in the other channel.}

To better demonstrate and justify the validity of our approach, we investigate two typical quantum conductors in the field of electron quantum optics: (1) a perfectly transparent quantum contact and (2) a mesoscopic capacitor.

\section{Perfectly transparent quantum contact}
\label{sec3}

We first concentrate on the case for perfectly transparent quantum contact, whose
many-body state has been studied by using the extended Keldysh-Green's function technique in previous works.\cite{vanevi2016, vanevi2017}

A quantum contact can be modeled as a quantum conductor, whose electron dwell time is negligible so that the instant
scattering approximation holds.\cite{beenakker2005,keeling2006,ivanov1997} Within the model we have presented in the previous section, a typical AC-biased quantum contact
can be described by the scattering matrix in the time-domain as
\begin{eqnarray}
  \left[
  \begin{tabular}{cc}
    $S_{\rm 11}(t)$& $S_{\rm 12}(t)$ \\
    $S_{\rm 21}(t)$& $S_{\rm 22}(t)$ \\            
  \end{tabular} \right] & = & \left(
                              \begin{tabular}{cc}
                                $\sqrt{D} e^{-i \phi(t)}$& $-i \sqrt{1-D}$ \\
                                $-i \sqrt{1-D}$& $\sqrt{D} e^{i \phi(t)}$ \\            
                              \end{tabular}
  \right),   
  \label{s3:eq1}
\end{eqnarray}
where $\phi(t) = \int^t d\tau V(\tau)$ is due to the contribution of the driven field $V(t)$, while $D$
characterizes the transparency of the contact. The scattering matrix in the energy domain[Eq.~\eqref{s2:eq1}] can be
obtained via the Fourier transform
\begin{eqnarray}
  S_{\rm \lambda \eta}(E, E') & = & \int dt e^{i (E-E') t} S_{\rm \lambda \eta}(t),
  \label{s3:eq2}
\end{eqnarray}
with $\lambda, \eta = 1, 2$. Note that within the instant scattering approximation, the scattering matrix $S_{\rm
  \lambda \eta}(E, E')$ is only the function of the energy difference $E-E'$.
  
In the full transparent limit $D \to 1$, the scattering between the two channels is absent and the many-body state can be written as the direct product of the two channels. Without loss of generality, we focus on channel $1$. Assuming a harmonic driving with $V(t) = V_0 \cos(\Omega_0 t)$, the corresponding Floquet scattering matrix elements for channel $1$ [Eq.~\eqref{s2:eq2}] can be given as
\begin{eqnarray}
  b_{\rm 1 n}(\delta) = \sum_m  S_{\rm nm} a_{\rm 1 m}(\delta),
  \label{s3:eq3}
\end{eqnarray}
where the element of the Floquet scattering matrix has the form: $S_{\rm nm} = J_{\rm n-m}(\xi)$ with $J_k(\xi)$ being the Bessel function of the first kind and $\xi = V_0/\Omega_0$ characterizing the strength of the driving field. Note that in this case, the matrix
element $S_{\rm nm}$ do not dependent on $\delta$, hence we shall omit it in the following discussion when no confusion can arise.

In the particle-hole basis, the Floquet scattering matrix $S$ can be decomposed into sub-matrices following
Eq.~\eqref{s2:eq5-2}, whose SVD can be done numerically using the LAPACK algorithm. Suppose the
SVD can be given as
\begin{eqnarray}
  S^{\rm pp} & = & U^p P [V^p]^{\dagger}, \nonumber\\
  S^{\rm hh} & = & U^h P [V^h]^{\dagger}, \nonumber\\
  S^{\rm ph} & = & U^p i \sqrt{ I - P^2 } [V^h]^{\dagger}, \nonumber\\
  S^{\rm hp} & = & U^h i \sqrt{ I - P^2 } [V^p]^{\dagger},
  \label{s3:eq4}
\end{eqnarray}
the corresponding many-body state of the outgoing electrons in channel $1$ can be written as
\begin{eqnarray}
  \hspace{-0.5cm}| \Psi_{\rm b1} \rangle & = & \prod_{\rm \delta \in [0, 1)}
                                               \prod_{k = 1, 2, ...} \Big[ \sqrt{ 1 - p_k}
                                               \nonumber\\
                                         && \hspace{1.5cm}\mbox{}+  i \sqrt{p_k}
                                            B^{\dagger}_{\rm p k}(\delta)
                                            B^{\dagger}_{\rm h k}(\delta) \Big] | F_{\rm L} \rangle,
                                            \label{s3:eq5}
\end{eqnarray}
with
\begin{eqnarray}
  B_{\rm p k}(\delta) & = & \sum_{\rm n \ge 0} [ U^{\rm p}_{\rm n k} ]^{\ast}
                                b^p_{\rm 1 n}(\delta), \nonumber\\
  B_{\rm h k}(\delta) & = & \sum_{\rm n < 0} U^{\rm h}_{\rm n k}
                                b^h_{\rm 1 n}(\delta).
                                \label{s3:eq6}
\end{eqnarray}
Note that although the scattering matrix does not depend on $\delta$, the creation(annihilation) operator for the particle and hole components is still $\delta$-dependent. Following previous works,\cite{vanevi2016, vanevi2017} we define the wave function of the particle and hole components ($U^{\rm p}_{\rm n k}$ and $U^{\rm h}_{\rm n k}$) in the time domain as
\begin{eqnarray}
  \psi^{\rm p}_k(t) & = & \sum_{\rm n \ge 0} e^{ - i n \Omega_0 t} U^{\rm p}_{\rm n k},
                          \nonumber\\
  \psi^{\rm h}_k(t) & = & \sum_{\rm n < 0} e^{ - i n \Omega_0 t} U^{\rm h}_{\rm n k}.
  \label{s3:eq7}
\end{eqnarray}

\begin{figure}
  \centering
  \includegraphics[width=6.5cm]{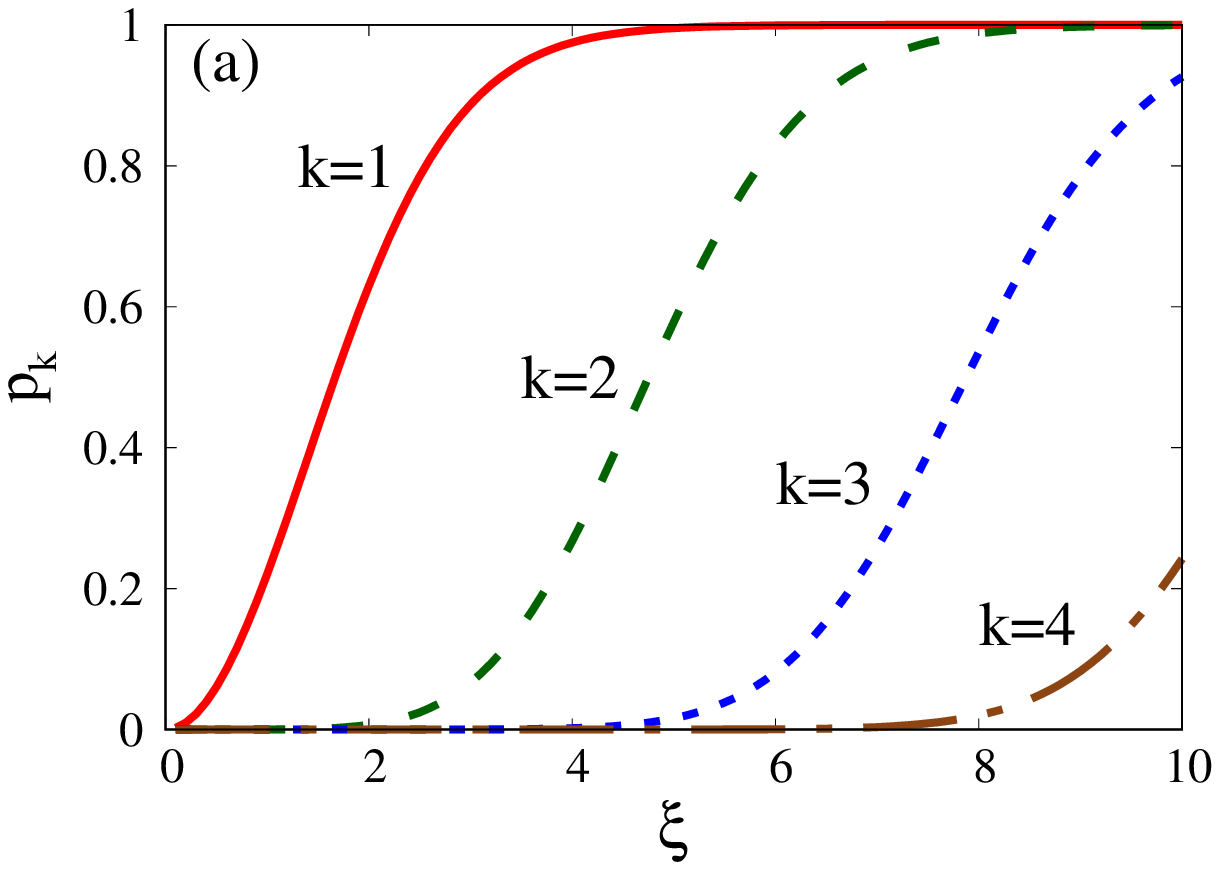}\\
  \includegraphics[width=6.5cm]{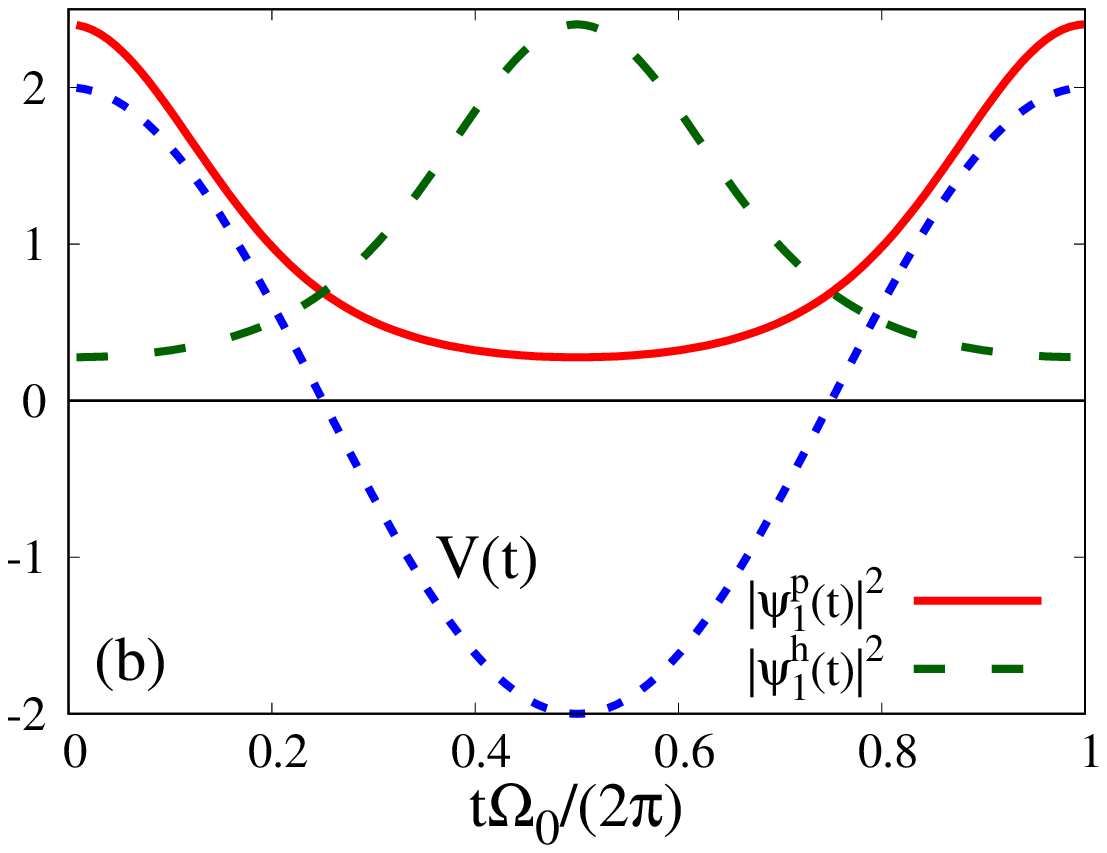}
  \caption{ (Color online) (a) The excitation probabilities $p_k$ of the first four modes($k=1,2,3,4$) as functions of the strength of the driving field $\xi$. Curves with different colors and line types correspond to different modes. (b) The particle $|\psi^p_1(t)|^2$(red solid curve) and hole $|\psi^h_1(t)|^2$(green dashed curve) components of the wave function for the first mode($k=1$) with the strength of driving $\xi = 2.0$. The driving field $V(t)$ is also plotted with blue dotter curve for comparison.}
  \label{fig2}
\end{figure}

The relation between the excitation probabilities $p_k$ and the strength of the driving field $\xi$ are given in
Fig.~\ref{fig2}(a), one can see that multiple modes of particle-hole pairs can be created during the scattering, while all the excitation probabilities $p_k$ are monotonically increasing functions of the strength $\xi$. The particle $|\psi^p_1(t)|^2$(red solid curve) and hole $|\psi^h_1(t)|^2$(green dashed curve) components of the wave function for the first mode $k=1$ are given in Fig.~\ref{fig2}(b), corresponding to the strength $\xi = 2.0$. We find that the particle and hole components of the wave function
($U^{\rm p}_{\rm n k}$ and $U^{\rm h}_{\rm n k}$) are related via
\begin{eqnarray}
  U^h_{\rm n,k} & = & i (-1)^{\rm n+1} U^p_{\rm -(n+1),k},
  \label{s3:eq8}
\end{eqnarray}
where $U^p_{\rm nk}$ is always real and $U^h_{\rm nk}$ is always imaginary. As a result,
the corresponding wave functions in time domain satisfies the relation
\begin{eqnarray}
  \psi^h_k(t) & = & \psi^p_k(t + \frac{\pi}{\Omega_0}).
  \label{s3:eq9}
\end{eqnarray}
Such relation indicates that the particle components are always in phase with the driving field, while the hole components are always 180 degrees out of phase. This can be understood as a direct consequence of the instant scattering approximation: since the dwell time of electrons in the quantum contact is negligible, the corresponding quasiparticles created during the scattering can response instantly to the driving field. This feature can be clearly seen in Fig.~\ref{fig2}(b) by comparing the profile of $|\psi^{\rm p(h)}_1(t)|^2$ with the driving field $V(t)$ in the time domain.

\begin{figure}
  \centering
  \includegraphics[width=6.5cm]{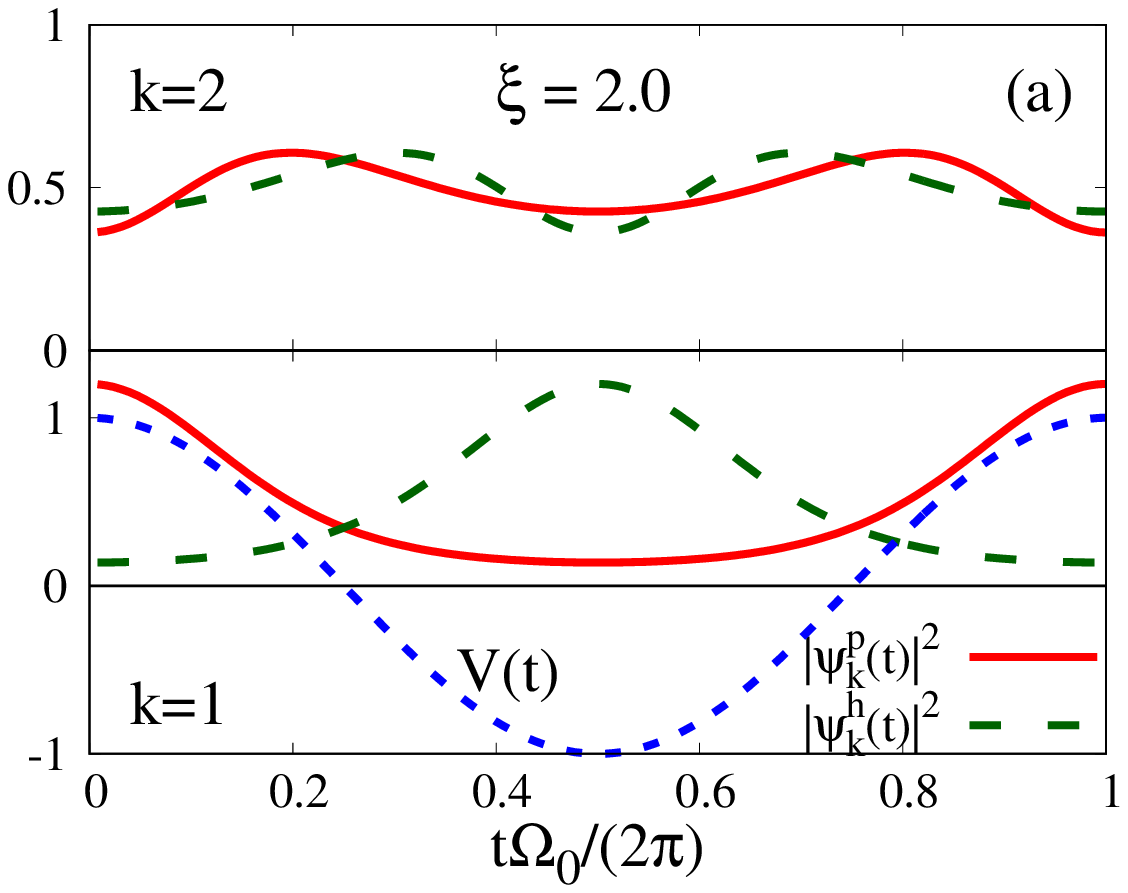}\\
  \includegraphics[width=6.5cm]{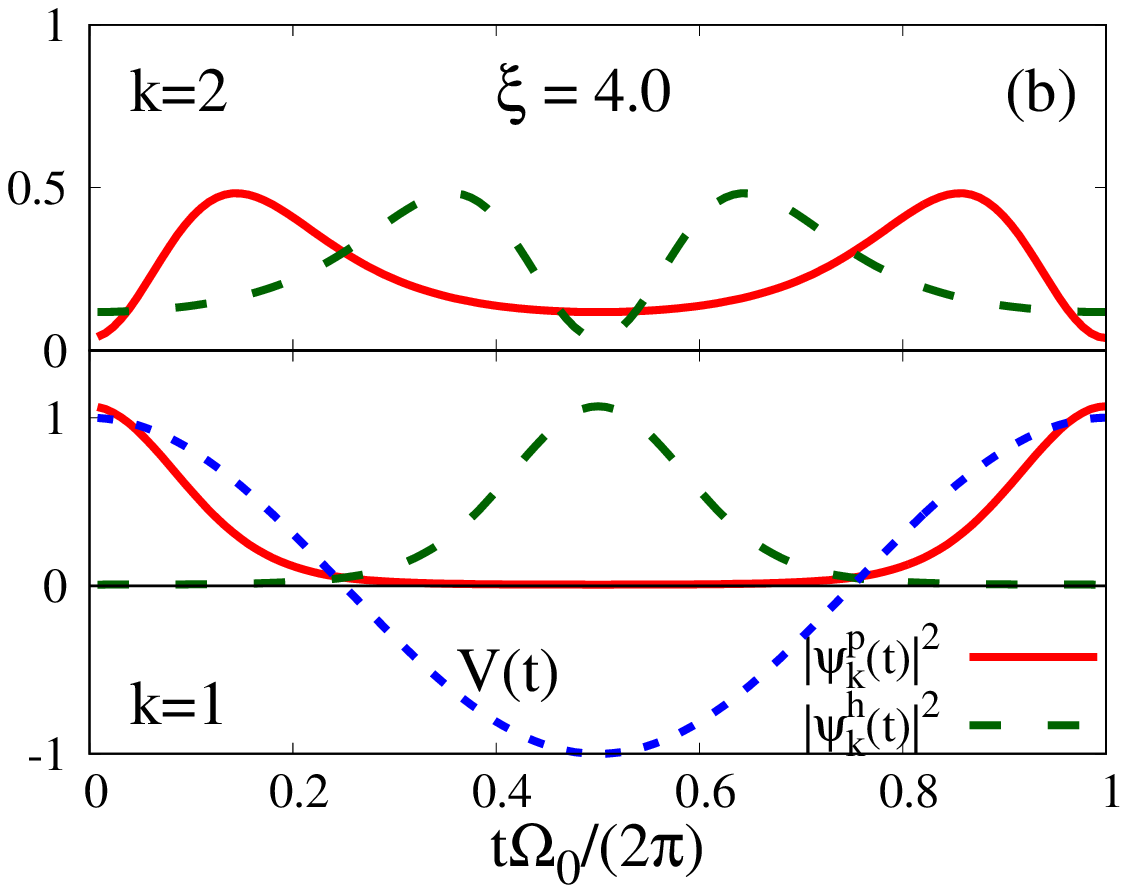}\\
  \includegraphics[width=6.5cm]{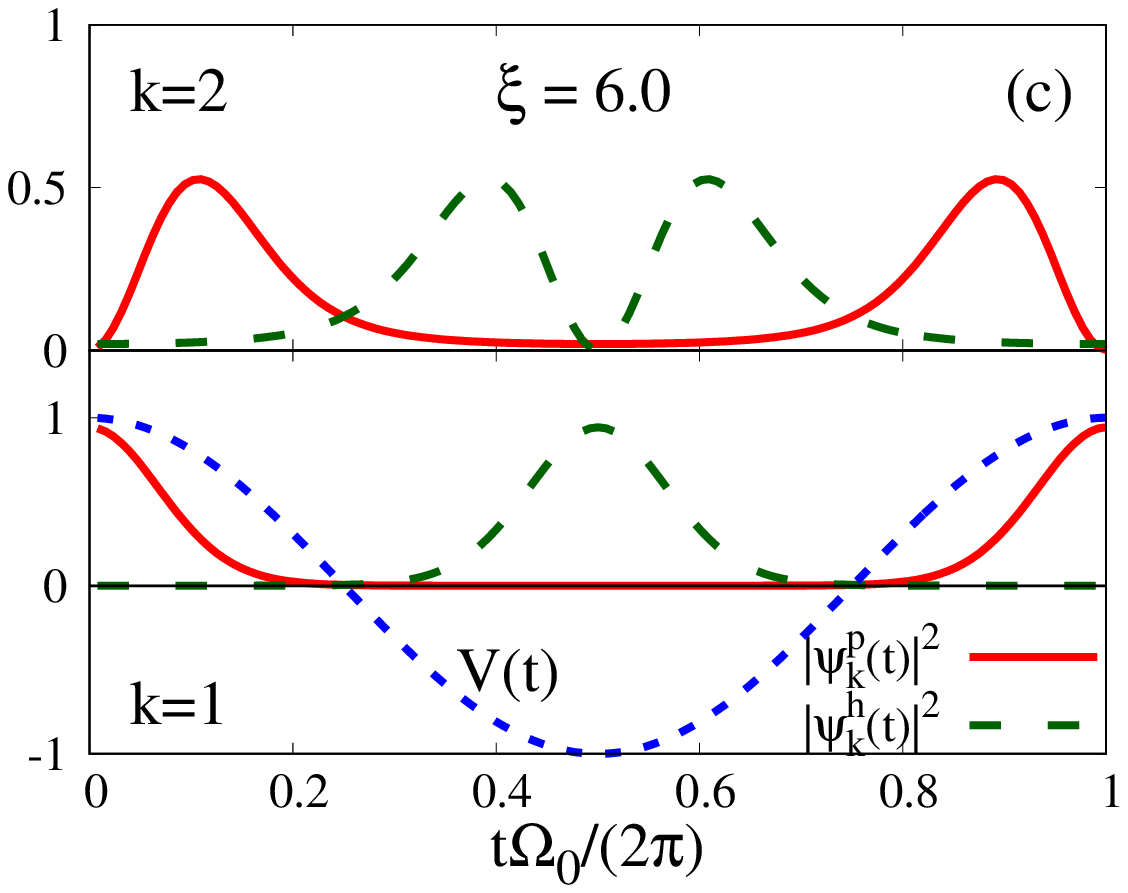}  
  \caption{ (Color online) The normalized particle $|\psi^p_k(t)|^2$(red solid curve) and hole $|\psi^h_k(t)|^2$(green dashed curve) components of the wave function for the first two modes($k=1, 2$) with (a) $\xi = 2.0$, (b) $\xi = 4.0$, and (c) $\xi = 6.0$. The driving field $V(t)$ is also plotted with blue dotter curve for comparison. Both the profile of $|\psi^{\rm p(h)}_k(t)|^2$ and the driving field $V(t)$ are normalized by the strength $\xi$.}
  \label{fig3}  
\end{figure}

Note that the relation given in Eqs.~\eqref{s3:eq8} and~\eqref{s3:eq9} hold for any value of $\xi$ and $k$. This is demonstrated in Fig.~\ref{fig3}, where the wave function of the first($k=1$) and the second($k=2$) modes are plotted with different value of the strength $\xi$. For relatively weak driving field[Fig.~\ref{fig3}(a) $\xi=2.0$], one can see that the particle(hole) component of the first mode exhibits a single peak, while the second ones has a double-peak structure. As the strength of the driving field $\xi$ increasing, the overall profile of the particle and hole components preserved, {\em i. e.}, no additional peak and/or valley is developed as $\xi$ increasing. In the meantime, the position of the peaks remains unchanged, indicating that the phase delay is unaffected by the value of $\xi$. The main impact of the strength $\xi$ is that, the particle(hole) components of the wave function tend to concentrate on the peak(valley) of the driving field in time domain. This makes the overlap between the particle and hole components become more and more smaller as the strength $\xi$ increasing. This can be seen most clearly for the first mode($k=1$), where only one peak exists in the profile of particle(hole) component. Hence, for strong driving field, the quantum contact tends to emit particles in the positive half cycle of the driving field, while tends to emit holes in the negative half cycle.

It is worth noting that by comparing the results in Fig.~\ref{fig2} with the ones given in Ref.~[\onlinecite{vanevi2017}], one can find that both the excitation probability and the wave function coincide, indicating that our approach agrees with the ones based on the extended Keldysh-Green's function technique, which has been used in Ref.~[\onlinecite{vanevi2017}].

\section{Mesoscopic capacitor}
\label{sec4}

As an on-demand single electron source, the mesoscopic capacitor plays an important role in the electron quantum
optics.\cite{feve2007} Besides the extensive studies on the average current and finite-frequency current
noise,\cite{buttiker1993, nigg2006, feve2007, moskalets2008, albert2010, mahe2010, lee2011} quantum tomographic methods
developed in recent years make it possible to extract the information of individual electronic
excitations.\cite{grenier2011a, bocquillon2013, jullien2014} In this section, we apply our approach to investigate the
wave function of quasiparticles created in a AC-driven mesoscopic capacitor. 

\begin{figure}
  \centering
  \includegraphics[width=6.5cm]{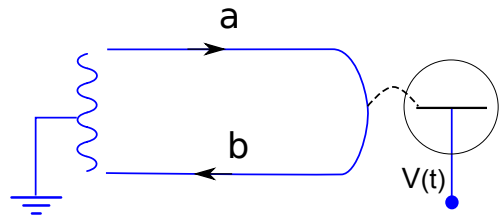}
  \caption{ (Color online) Schematic of a mesoscopic capacitor, composed of a single-level QD coupled to a quantum hall edge channel. The QD is driven by a time-dependent voltage $V(t)$.}
  \label{fig4}
\end{figure}

In a simple case, the mesoscopic capacitor can be modeled as a single-level quantum dot (QD) coupled to a quantum hall edge channel via a quantum point contact, as illustrated in Fig.~\ref{fig4}. By driving the QD via a AC voltage $V(t)$, the mesoscopic capacitor can emit(absorb) electrons to(from) the edge channel, making it capable to serve as a coherent electron source. Here we focus on the case for harmonic driving with $V(t) = V_0 \cos(\Omega_0 t)$.

The incoming and outgoing electrons along the edge channel can be related via the scattering matrix as
\begin{eqnarray}
  b(E) & = & \int \frac{dE'}{2\pi} S(E, E') a(E'),
  \label{s4:eq1}
\end{eqnarray}
with
\begin{eqnarray}
  S(E, E') & = & \int dt e^{i E t} \int dt' e^{-i E t'} S(t, t').
  \label{s4:eq2}
\end{eqnarray}
The scattering matrix in the time domain can be obtained following Ref.~[\onlinecite{keeling2008}], which has the form
\begin{eqnarray}
  S(t, t') & = & \delta(t-t') - 2 \Gamma \Theta(t-t') e^{-\Gamma (t-t')} e^{-i \epsilon_0 (t - t')} \nonumber\\
           && \times e^{-i[\phi(t) - \phi(t')]},
  \label{s4:eq3}
\end{eqnarray}
with $\Gamma$ characterizing the coupling between the QD and the edge channel, while $\epsilon_0$ representing the
energy level of the QD. The phase $\phi(t) = \int^t d\tau V(\tau)$ is due to the time-dependent voltage applied on the QD. The notation $\Theta(t)$ represents the Heaviside step function. Note that due to the finite dwell time of the electron in the QD (dominated by the QD-channel coupling $\Gamma$), the instant scattering approximation does not hold for such system. 

The corresponding Floquet scattering matrix[Eq.~\eqref{s2:eq2-1}] can be calculated from Eqs.~\eqref{s4:eq2} and~\eqref{s4:eq3}, which has the form
\begin{eqnarray}
  && S_{\rm nm}(\delta) = \delta_{\rm nm} \nonumber\\
  && \hspace{0.5cm}\mbox{}+ 2 \Gamma \sum_k \frac{ J_k(\xi) J_{\rm m - n + k}(\xi) }{i\Omega_0[(n-k) + \delta - \epsilon_0/\Omega_0 ] - \Gamma},
  \label{s4:eq4}
\end{eqnarray}
with $\xi = V_0/\Omega_0$ characterizing the strength of the driving field. Note that in this case, the Floquet scattering matrix has an  $\delta$-dependence, which is different from the case for quantum contact.

The many-body state of the system can be obtained following the same procedure described in the previous section, leading to the form
\begin{eqnarray}
  \hspace{-0.5cm}| \Psi_{\rm b} \rangle & = & \prod_{\rm \delta \in [0, 1)}
                                               \Big[ \sqrt{ 1 - p(\delta)}
                                               \nonumber\\
                                         && \hspace{1.5cm}\mbox{}+  i \sqrt{p(\delta)}
                                            B^{\dagger}_{\rm p}(\delta)
                                            B^{\dagger}_{\rm h}(\delta) \Big] | F \rangle,
                                            \label{s4:eq5}
\end{eqnarray}
with $| F \rangle$ representing the Fermi sea of the edge channel, while the particle and hole components can be given as
\begin{eqnarray}
  B_{\rm p}(\delta) & = & \sum_n [ U^{\rm p}_n(\delta) ]^{\ast} b^p_n(\delta), \nonumber\\
  B_{\rm h}(\delta) & = & \sum_n U^{\rm h}_n (\delta) b^h_n(\delta).
                                \label{s4:eq6}
\end{eqnarray}
In a similar way, we can also define the particle and hole component of the wave function in time-domain as
\begin{eqnarray}
  \psi^{\rm p}(\delta, t) & = & \sum_{\rm n \ge 0} e^{ - i n \Omega_0 t} U^{\rm p}_n(\delta),
                          \nonumber\\
  \psi^{\rm h}(\delta, t) & = & \sum_{\rm n < 0} e^{ - i n \Omega_0 t} U^{\rm h}_n(\delta).
  \label{s4:eq6-1}
\end{eqnarray}

Note that there are two main differences between this case and the case of the quantum contact [Eqs.~\eqref{s3:eq5} and~\eqref{s3:eq6}]. One is that, there is only one mode with nonzero excitation probability $p(\delta)$, {\em i. e.}, the mesoscopic capacitor can create only one mode the of particle-hole pair in the Fermi sea of the edge channel, despite the strength of the driving field. The other one is that, the excitation probability $p(\delta)$ and the particle[hole] component of the wave function $U^{\rm p}_n(\delta)$[$U^{\rm h}_n(\delta)$] are the function of $\delta$. This is a direct consequence of the break down of the instant scattering approximation,\cite{beenakker2005} leading to additional energy dependence of the scattering matrix.\cite{vavilov2001} 

\begin{figure}[h]
  \centering
  \includegraphics[width=7.5cm]{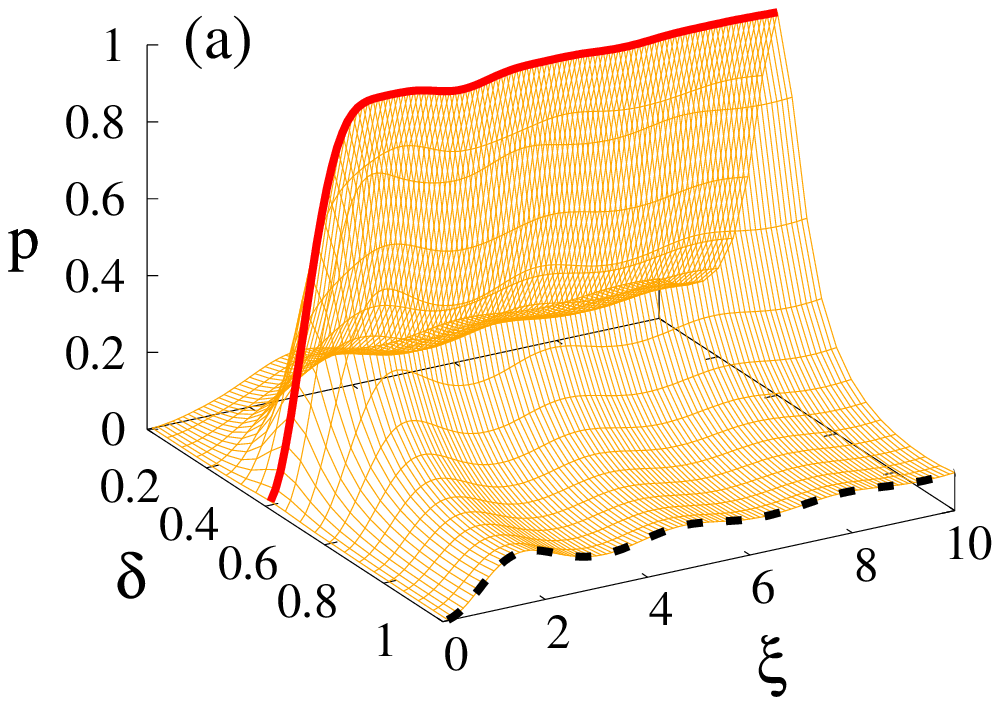}\\
  \includegraphics[width=6.5cm]{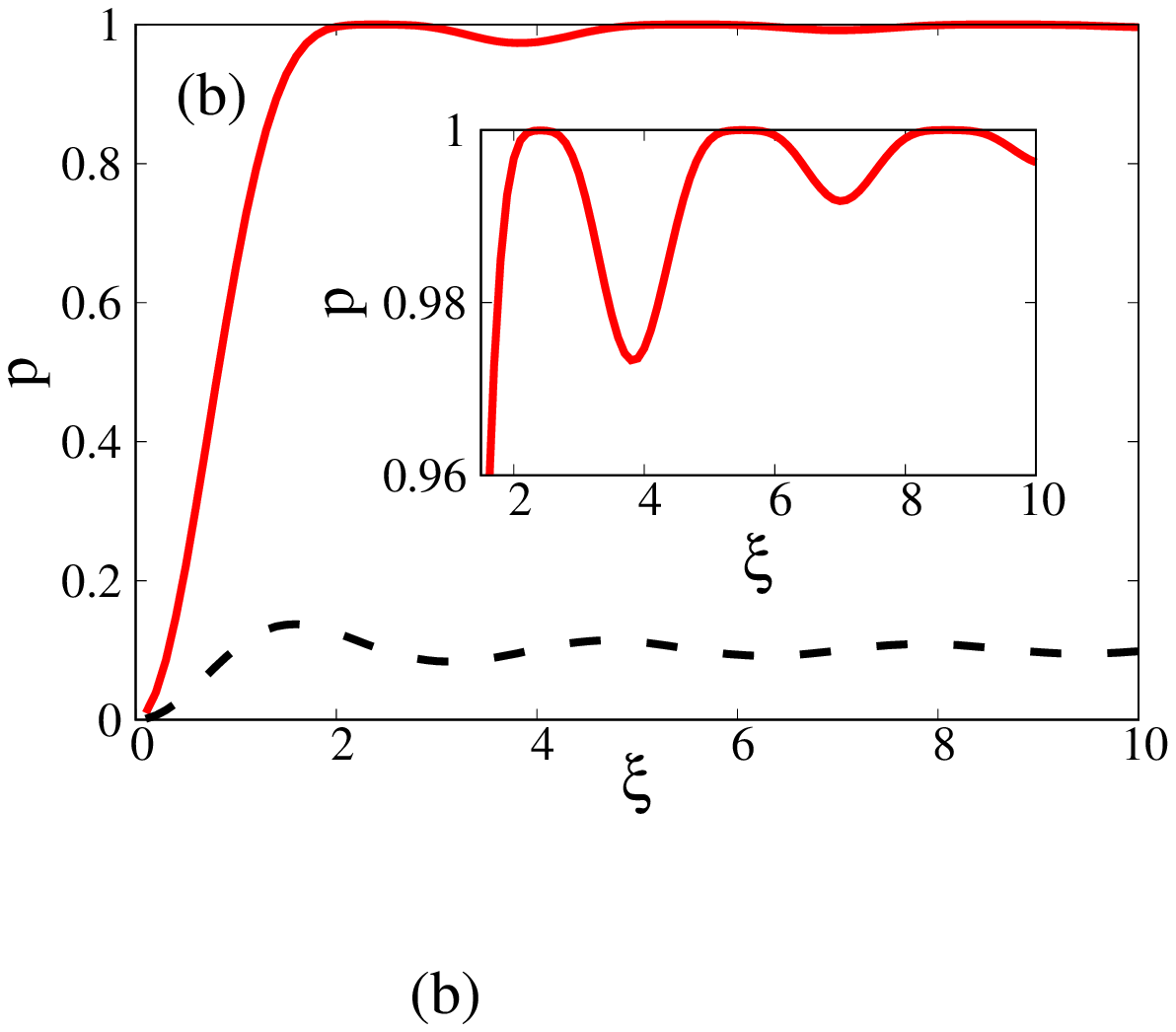}\\
  \includegraphics[width=7.5cm]{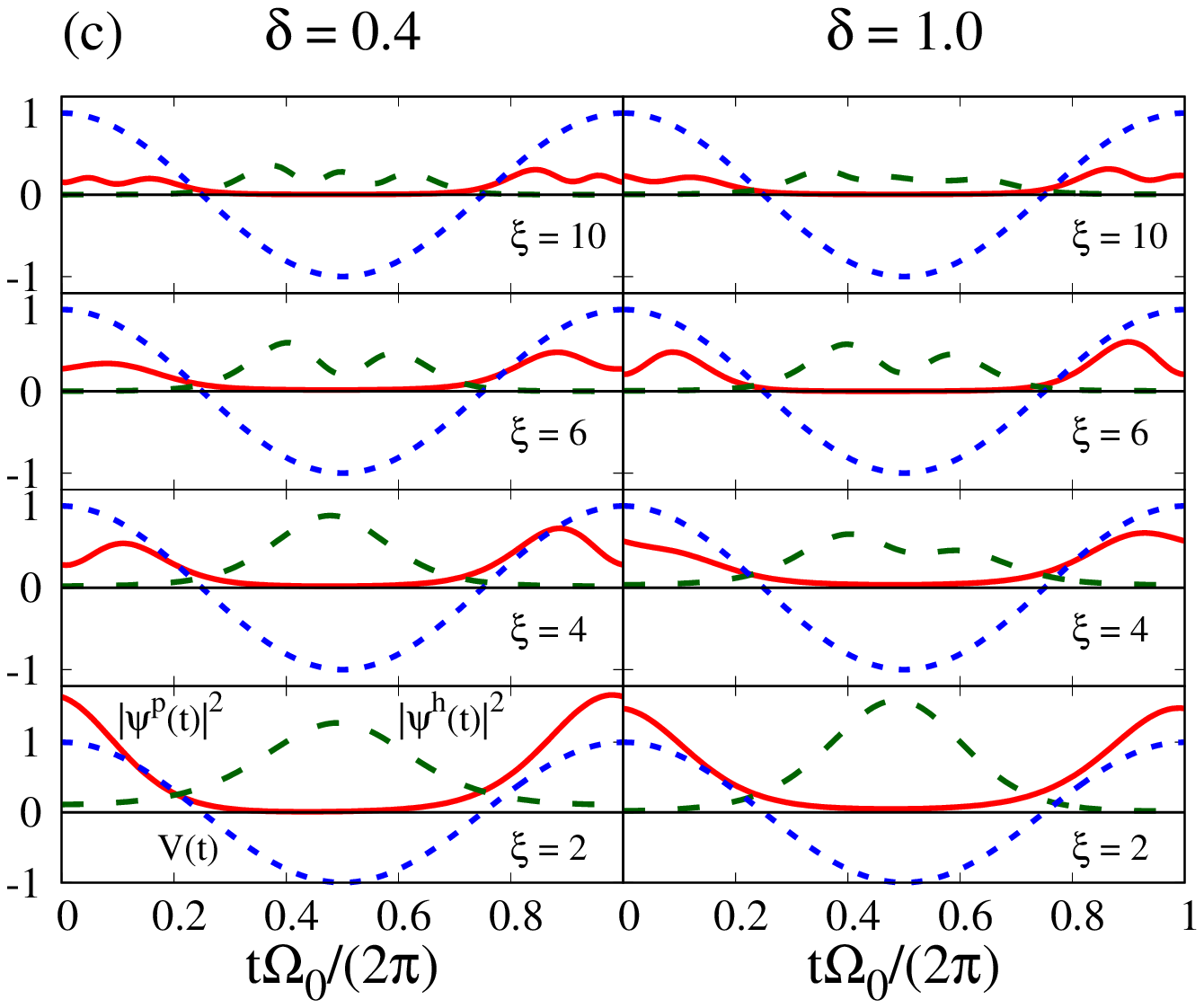}
  \caption{ (Color online) (a) The dependence of the excitation probability $p$ on the parameter $\delta$ and the strength of driving $\xi$. The red solid and black dotted curves corresponds to the case for $\delta=0.4$ and $\delta=1.0$, respectively. (b) The dependence of the excitation probability $p$ on the strength of driving $\xi$ for $\delta=0.4$(red solid curves) and $\delta=1.0$(black dotted curves), which exhibits oscillations. The details of the oscillation for $\delta=0.4$ is shown in the inset. (c) The normalized profile of the particle $|\psi^h(t)|^2$ (red solid curves) and hole $|\psi^h(t)|^2$(red solid curves) components of the wave function for different value of $\xi$ and $\delta$. The driving field $V(t)$ is also plotted with blue dotter curve for comparison. Both the profile of $|\psi^{\rm p(h)}(t)|^2$ and the driving field $V(t)$ are normalized by the strength $\xi$.}
  \label{fig5}
\end{figure}

Let us first concentrate on the excitation probability $p$, which is both the function of $\delta$ and $\xi$. According to the structure of the denominator in Eq.~\eqref{s4:eq4}, a resonant peak is expected under the resonant condition $\delta =  \epsilon_0/\Omega_0$. We find that such resonant peak can be clearly seen for small $\Gamma$, as shown in Fig.~\ref{fig5}(a). In the figure, we have chosen $\Gamma/\Omega_0=0.1$ and $\epsilon_0/\Omega_0=0.4$. One can see that for a given strength of driving $\xi$, the excitation probability $p$ is peaked around $\delta = 0.4$, which agrees with the resonant condition.

The dependence of the probability $p$ on the strength of driving $\xi$ is more interesting. From Fig.~\ref{fig5}(a), an oscillation structure can be identified. To better demonstrate this, we plot the $\xi$-dependence of $p$ at the resonance ($\delta = 0.4$) and far away from the resonance ($\delta =1.0$) in Fig.~\ref{fig5}(b). From the figure, one can see that for $\delta = 0.4$(red solid curve), the probability $p$ monotonically increases and almost reaches $1$ as $\xi$ increasing from $0$ to $2.42$. Then, as $\xi$ further increasing, $p$ undergoes an oscillation, resulting in 3 peaks and valleys for $\xi < 10$. A similar oscillation behavior can also be seen for $p$ with $\delta = 1.0$.

Now we turn to the particle and hole components of the wave functions. In Fig.~\ref{fig5}(c), we compare the profile of the particle[$|\psi^p(t)|^2$, red solid curves] and hole [$|\psi^h(t)|^2$, green dashed curves] components at the resonance($\delta = 0.4$, left column) and away from the resonance ($\delta =1.0$, right column) for different strength of driving $\xi$. One can see that, for a given strength $\xi$, $|\psi^{p(h)}(t)|^2$ is not sensitive to the value of $\delta$, which can be seen by comparing the curves in the left column to the corresponding curves in the right column. In contrast, their dependence on the strength $\xi$ are more pronounced. By increasing $\xi$, the profile of $|\psi^{p(h)}(t)|^2$ can develop additional peaks as $\xi$ increasing. Note that this is different from the case of the quantum contact, where the profile of  the particle and hole components in the time domain do not change as the strength $\xi$ increasing.

Also note that unlike the case for quantum contacts, both the particle($U^p_n$) and hole($U^h_n$) components has nonzero real and imaginary part, indicating that a phase delay exists between the driving field and the particle-hole pair. Since the dwell time of the electron in the QD of the mesoscopic capacitor is decided by the QD-channel coupling $\Gamma$, we can attributed such phase delay to the contribution of the finite dwell time of the electron in the quantum conductor. The phase delay can be better seen for the cases with $\xi=2$ where only one peak is developed in the profiles of the wave function[the bottom row in Fig.~\ref{fig5}(c)]: The overlap between the profile of $|\psi^{\rm p}(t)|^2$(red solid curve) and the driving field $V(t)$ indicates that they are not exactly in phase.

\begin{figure}[h]
  \centering
  \includegraphics[width=7.5cm]{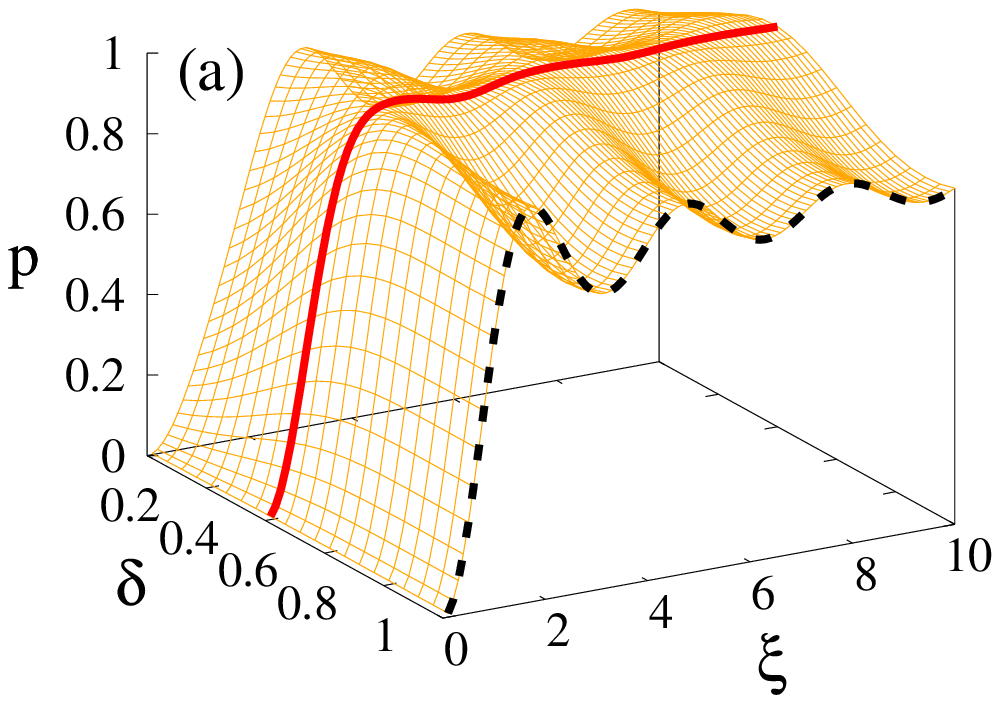}\\
  \includegraphics[width=6.5cm]{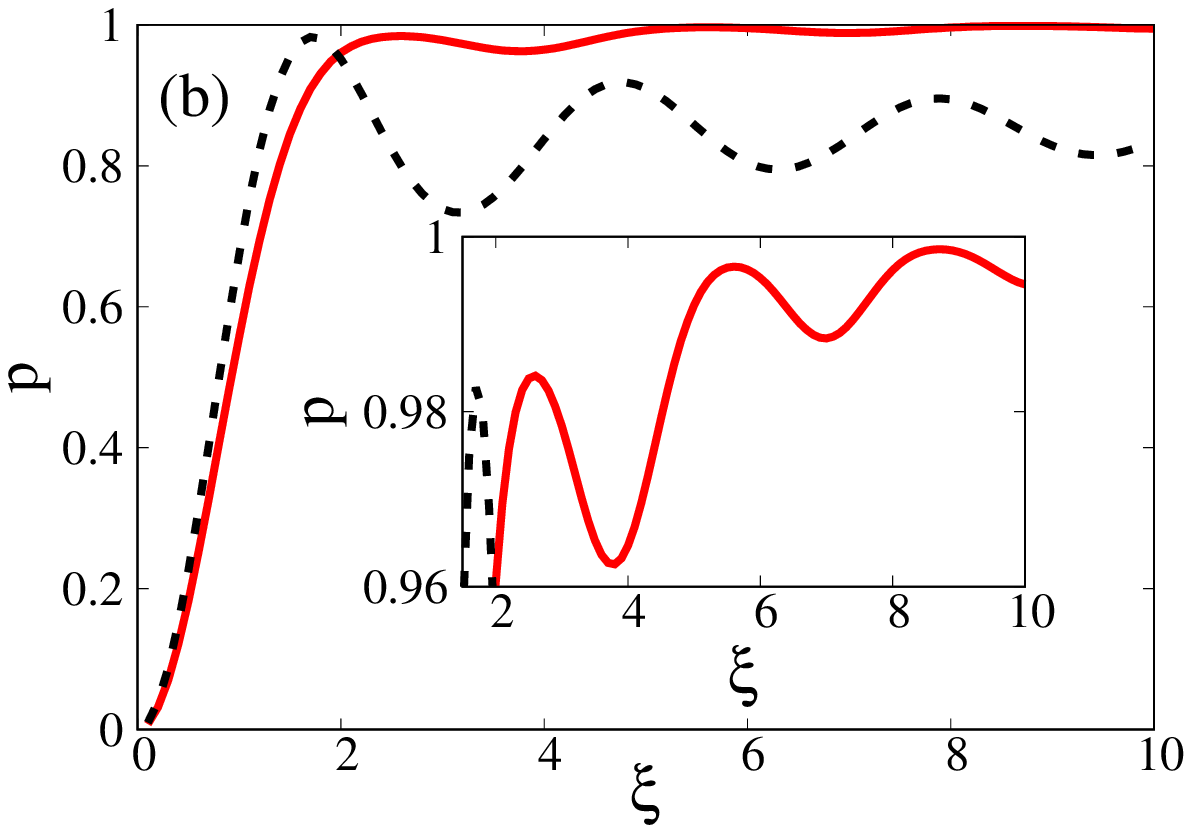}\\
  \includegraphics[width=7.5cm]{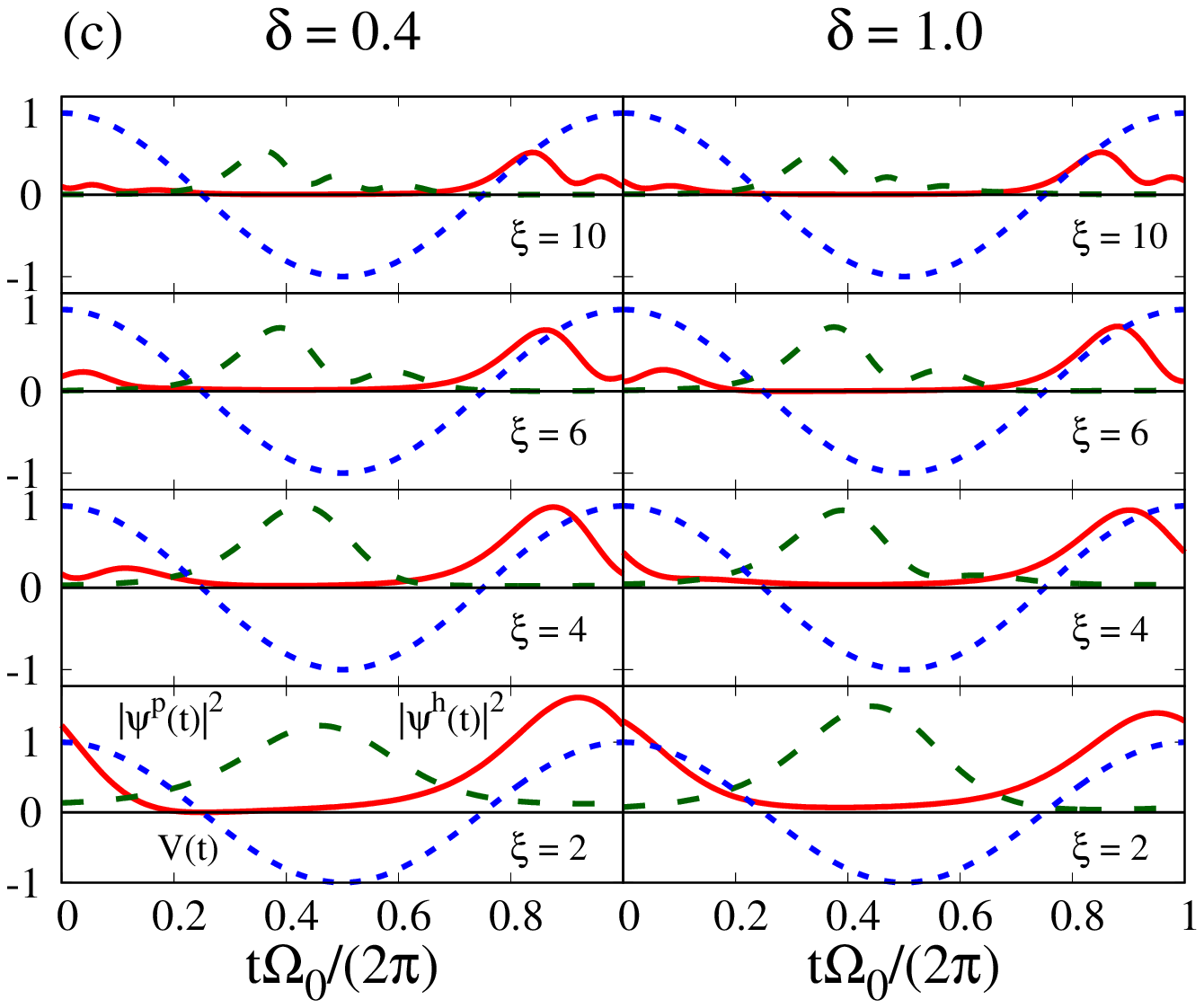}
  \caption{ (Color online) The same as Fig.~\ref{fig5}, but with the QD-channel coupling $\Gamma=0.5$.}
  \label{fig6}
\end{figure}

By increasing the QD-channel coupling $\Gamma$, both the oscillation of the excitation probability $p$ and the phase delay of the wave function can be altered. To show this, we plot the probability $p$ and the corresponding $|\psi^{\rm p(h)}(t)|^2$ for $\Gamma=0.5$ in Fig.~\ref{fig6}. By comparing Fig.~\ref{fig6}(a) with Fig.~\ref{fig5}(a), one can see that the probability $p$ away from the resonance is dramatically enhanced, leading to a broaden of the resonance peak. Such enhancement can be better seen in Fig.~\ref{fig6}(b), where the probability $p$ at the resonance($\delta = 0.4$) are comopared to the ones away from the resonance ($\delta =1.0$). In fact, the probability $p$ at the resonance is slightly suppressed for large $\Gamma$, which can be better seen from the inset of Fig.~\ref{fig6}(b).

The phase delay between the driving field and the particle(hole) components $|\psi^{\rm p(h)}(t)|^2$ is increased by increasing the QD-channel coupling $\Gamma$. This can be demonstrated by comparing their profiles with the driving field $V(t)$ in time domain, as shown in Fig.~\ref{fig6}(c). As in Fig.~\ref{fig5}(c), the increasing of the phase delay can be better seen for the cases with $\xi=2$[the bottom row of Fig.~\ref{fig6}(c)], where only a single peak exists for $|\psi^{\rm p(h)}(t)|^2$. By comparing to the corresponding figures in Fig.~\ref{fig5}(c), one can see that the overlap between the profile of $|\psi^{\rm p}(t)|^2$(red solid curve) and the driving field $V(t)$ are enlarged for large $\Gamma$ in Fig.~\ref{fig6}(c), indicating an increasing of the phase delay between them.

For larger $\xi$, additional peaks are developed, while the phase delay of these peaks has a different dependence on the QD-channel coupling $\Gamma$. The phase delay for peaks in the region with larger time $t$ tends to have a larger phase delay as $\Gamma$ increasing. As a consequence, the peaks tend to merge and concentrate on the region for small $t$, as can be seen by comparing the corresponding profile of $|\psi^{\rm p(h)}(t)|^2$ in Fig.~\ref{fig5}(c) and Fig.~\ref{fig6}(c).

\begin{figure}[h]
  \centering
  \includegraphics[width=7.5cm]{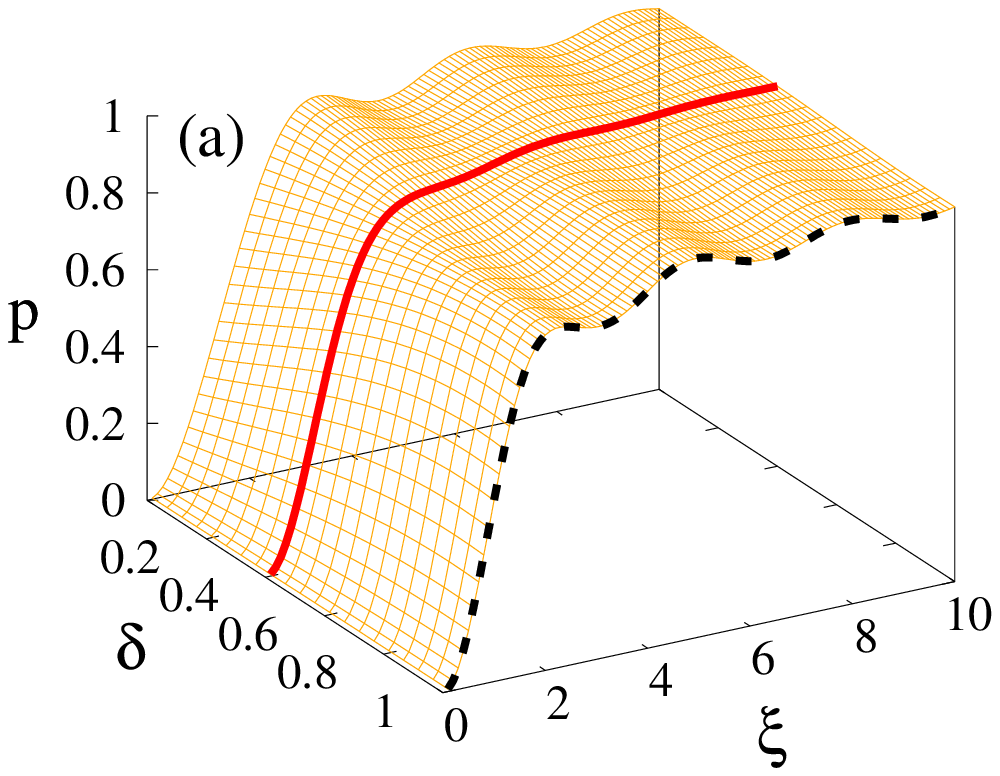}\\
  \includegraphics[width=6.5cm]{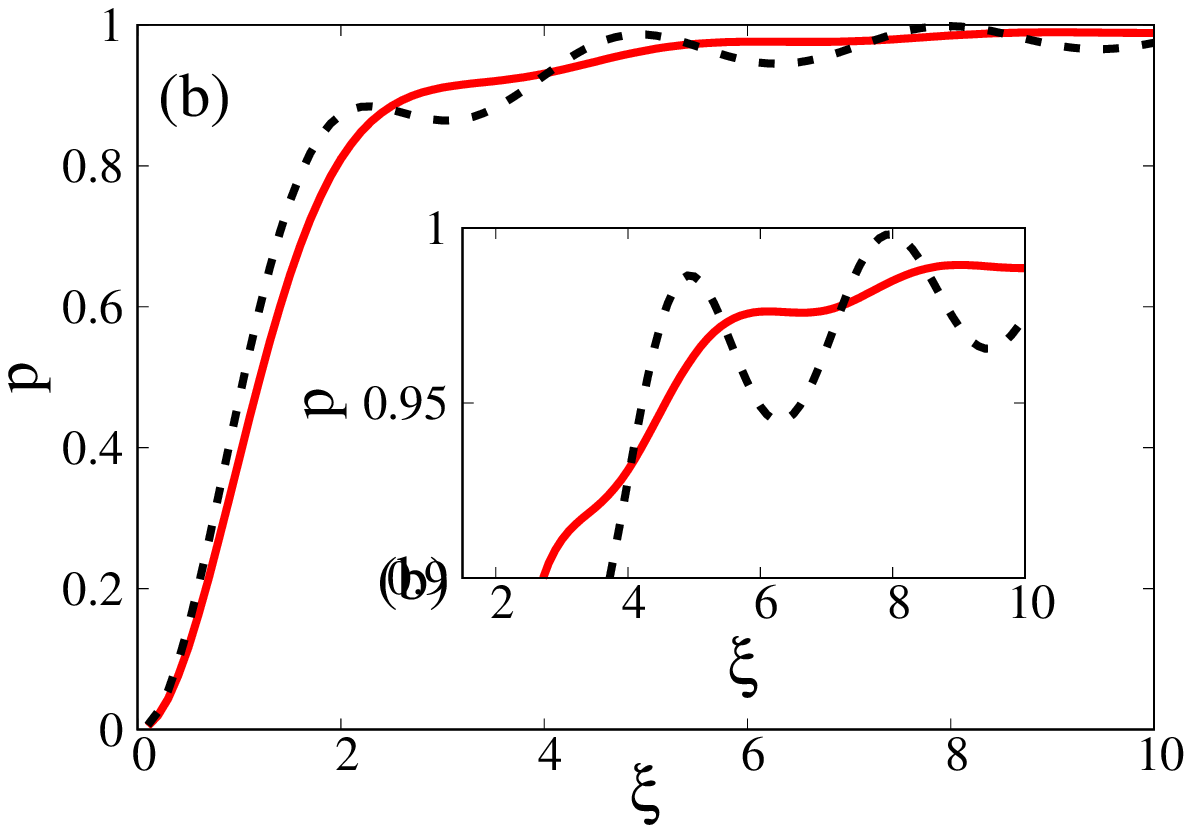}\\
  \includegraphics[width=7.5cm]{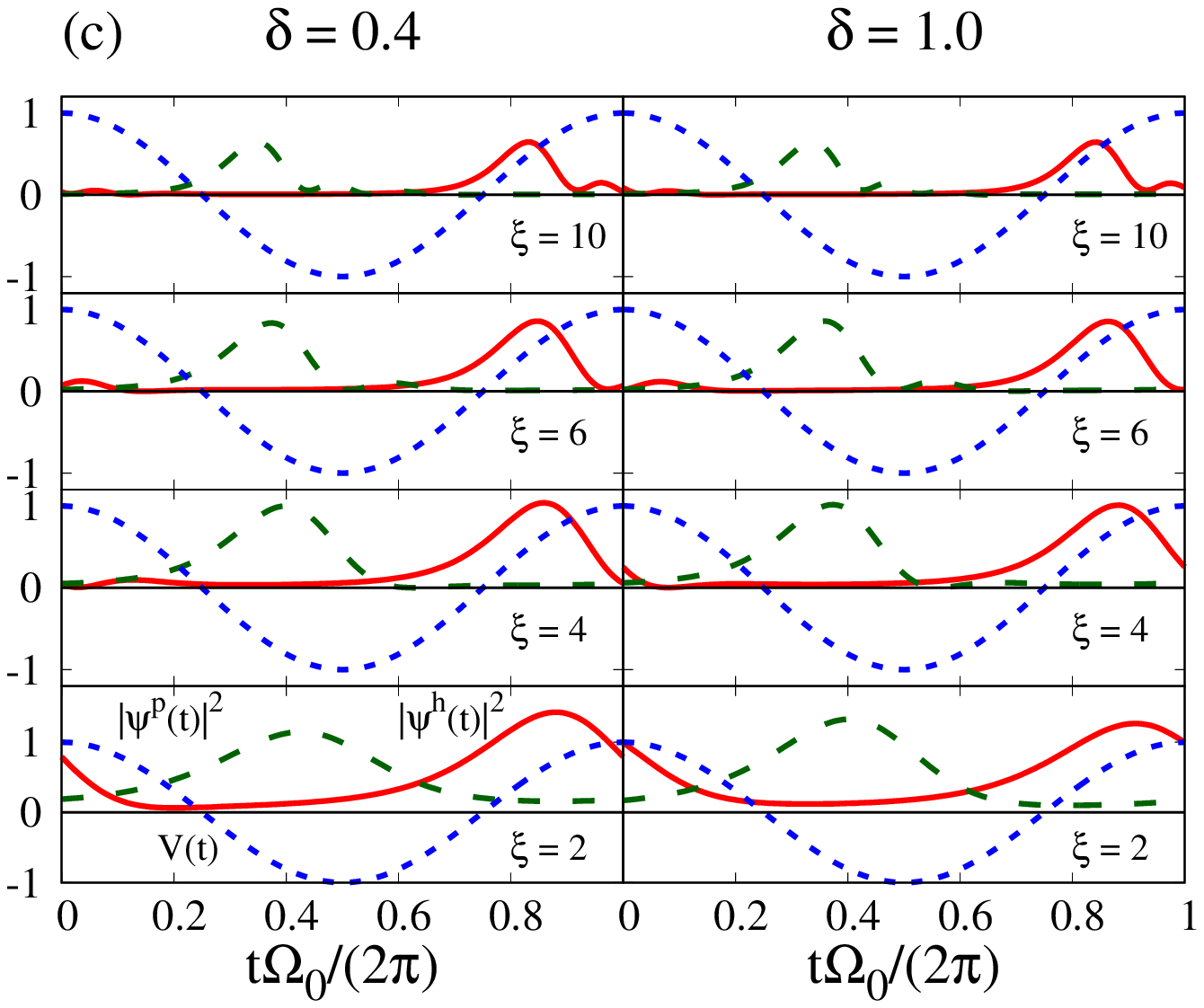}
  \caption{ (Color online) The same as Fig.~\ref{fig5}, but with the QD-channel coupling $\Gamma=1.0$.}
  \label{fig7}
\end{figure}

By further increasing $\Gamma$, both the oscillation and the resonance peak of the probability $p$ are smeared out. As a result, the probability $p$ becomes insensitive to $\delta$, while its dependence on the strength $\xi$ tends to be monotonous, as can be seen in Fig.~\ref{fig7}(a) and (b), which is calculated for $\Gamma = 1.0$. Accordingly, the phase delay of the wave function is further enhanced, which is shown in Fig.~\ref{fig7}(c). Note that in this case, due to the large difference of the phase delay between the peaks in the profile of $|\psi^{\rm p(h)}(t)|^2$, these peaks are merged almost into a single peak, as can be seen by comparing Fig.~\ref{fig7}(c) to Fig.~\ref{fig5}(c) and Fig.~\ref{fig6}(c).

\section{Relation to the first-order electronic correlation function}
\label{sec5}

In previous studies on the electron quantum optics, the first-order correlation function, or equivalently, the Wigner function, have been extensively used in the description of the first-order quantum coherence, which can be obtained via quantum tomographic methods.\cite{ferraro2013, bocquillon2014, jullien2014, grenier2011a} In a recent work, Marguerite {\em et al.} suggest a signal processing technique to extract the particle and hole wave functions of quasiparticles from such correlation functions in a quantum contact, which works in the case when only one particle-hole pair is relevant.\cite{marguerite2017} In this section, we show that our approach offers a systematic way to extract such wave functions from the first-order correlation function, which closely related to their work. For clarification, we demonstrate this in the single-channel case.

The information of quasiparticles created during the scattering is encoded in the first-order correlation function of the outgoing electrons. Following the notion introduced in Sec.~\ref{sec2}, it has the form in zero-temperature limit as
\begin{eqnarray}
G(t, t') & = & \langle \Psi_b | b^{\dagger}(t) b(t') | \Psi_b \rangle,
\label{s5:eq1}
\end{eqnarray}
where $b(t)$ representing the annihilation operator for outgoing electron in time domain. For driving field with fundamental frequency $\Omega_0$, $b(t)$ can be written in the Floquet space as
\begin{eqnarray}
b(t) = \sum_{\rm n = -\infty}^{+\infty} \frac{\Omega_0}{2\pi} \int^1_0  d\delta  e^{-i\Omega_0 (n + \delta) t} b_n(\delta).
\label{s5:eq2}
\end{eqnarray}
Accordingly, the first-order correlation function in the Floquet space has the form
\begin{eqnarray}
\hspace{-0.5cm}G(t, t') & = &  \sum_{\rm n,n' = -\infty}^{+\infty} e^{i \Omega_0 (n - n')(t + t')/2} g_{\rm \frac{n+n'}{2}}(t - t'),
\label{s5:eq3}
\end{eqnarray}
with
\begin{eqnarray}
g_l(\tau) & = & \int^1_0 d\delta e^{i (\delta + l) \Omega_0 \tau} \langle \Psi_b | b^{\dagger}_n(\delta) b_{n'}(\delta) | \Psi_b \rangle.
\label{s5:eq4}
\end{eqnarray}

The expectation over the outgoing state $| \Psi_b \rangle$ in the above equations can be calculated by using the quasiparticle representation given in Eqs.~\eqref{s2:eq11} and~\eqref{s2:eq12}, within which the first-order correlation function can be decomposed into two parts 
\begin{eqnarray}
G(t, t') & = & G_0(t, t') + \Delta G(t, t').
\label{s5:eq5}
\end{eqnarray}
The first part $G_0(t, t') = \langle F | b^{\dagger}(t) b(t') | F \rangle$ corresponds to the contribution of the Fermi sea,  which is irrelevant to the quasiparticle excitations. It is the second part $\Delta G(t, t')$, which has been referred as the excess single electron coherence, contains the information of the quasiparticles. It has the form (see Appendix~\ref{app3} for details)
\begin{eqnarray}
\Delta G(t, t') & = & \sum_{\rm k = 1, 2, ...} \int^1_0 d\delta \Big([\psi^p_k(\delta, t)]^{\ast}, [\psi^h_k(\delta, t)]^{\ast}\Big) \label{s5:eq6}\\
&& \hspace{-2.0cm} \times \left(\begin{tabular}{cc}
             $p_k(\delta)$ & $-i \sqrt{p_k(\delta)[1 - p_k(\delta)]}$\\
             $i \sqrt{p_k(\delta)[1 - p_k(\delta)]}$ & $ - p_k(\delta)$\\
           \end{tabular}\right) \left(\begin{tabular}{c}
             $\psi^p_k(\delta, t')$\\
             $\psi^h_k(\delta, t')$\\
           \end{tabular}\right), \nonumber
\end{eqnarray}
where we have defined the particle and hole components of the wave function in time domain as
\begin{eqnarray}
\psi^p_k(\delta, t) & = & \sum_{\rm n \ge 0} e^{-i \Omega_0 (\delta + n) t} U^p_{\rm nk}(\delta), \nonumber\\
\psi^h_k(\delta, t) & = & \sum_{\rm n < 0} e^{-i \Omega_0 (\delta + n) t} U^h_{\rm nk}(\delta).
\label{s5:eq7}
\end{eqnarray} 

The above equation indicates that the particle and hole components can be obtained by expressing the first-order correlation function in a "diagonalized" form. In fact, a simplified version of such procedure has already been suggested by Marguerite {\em et al.} in Ref.~[\onlinecite{marguerite2017}] for the case of the quantum contacts. In this case, the scattering matrix in the Floquet space is independent on the parameter $\delta$ as we have shown in Sec.~\ref{sec3}. The excess single electron coherence $\Delta G(t, t')$ can be simplified to
\begin{eqnarray}
\Delta G(t, t') & = & \sum_{\rm k = 1, 2, ...}\Big([\psi^p_k(t)]^{\ast}, [\psi^h_k(t)]^{\ast}\Big) \nonumber\\
&& \hspace{-1.75cm} \times \left(\begin{tabular}{cc}
             $p_k$ & $-i \sqrt{p_k[1 - p_k]}$\\
             $i \sqrt{p_k[1 - p_k]}$ & $ - p_k$\\
           \end{tabular}\right) \left(\begin{tabular}{c}
             $\psi^p_k(t')$\\
             $\psi^h_k(t')$\\
           \end{tabular}\right), \nonumber\\
          && \mbox{} \label{s5:eq8}
\end{eqnarray}
where we have chosen $t-t' = m \Omega_0$ with $m$ being an integer. This is just the "diagonalized" form suggested in Ref.~[\onlinecite{marguerite2017}], while the wave function $\psi^{\rm p(h)}_k(t)$ serves as the "electronic atoms of signal".\cite{fletcher2013} Hence the Eq.~\eqref{s5:eq6} can be regarded as a generalized version of their procedure, which works for a general quantum conductor.

\section{SUMMARY}
\label{sec6}

In this work, we have provide a general approach to extract the wave function of quasiparticles in AC-driven quantum conductors. This is done by incorporating the Bloch-Messiah reduction into the scattering theory approach to quantum transport. In doing so, we explicitly construct the many-body state from the scattering matrix, from which the wave function of the quasiparticles can be extracted. We have found that two species of quasiparticles can be excited in the system. Due to the electron number conservation, the quasiparticles have to be excited in pairs, and hence the elementary excitations created by the AC driving field are particle-hole pairs. 

By using our approach, we have compared the particle-hole pairs created in the perfectly quantum contact and the mesoscopic capacitor with single-level quantum dot. For the quantum contact, multiple modes of particle-hole pairs can be excited during the scattering, while the excitation probability of each pair has a monotonous dependence on the strength of the driving field. The particle(hole) components of the wave functions are alway in phase(180 degrees out of phase) with the driving field. These results agree with the ones obtained via the extended Keldysh-Green's function technique. For the mesoscopic capacitor, only a single mode of particle-hole pairs can be excited, while the corresponding excitation probability can undergo an oscillation as the strength of the driving field increasing. A phase delay between the particle(hole) components of the wave functions and the driving field exists, which can be attributed to the finite dwell time of the electron in the QD of the mesoscopic capacitor. 

Besides these theoretically studies, we have also find that our approach can also offer a procedure to extract the wave function of the quasiparticles from the first-order electronic correlation function, which can be helpful in the signal processing of quantum tomographic experiments. We expect the approach we have reported here to be helpful in the further study of single electronic excitations in electron quantum optics.

\begin{acknowledgments}
The author would like to thank Professor M. V. Moskalets for helpful comments and discussion. This work was supported by Key Program of National Natural Science Foundation of China under Grant No. 11234009, National Key Basic Research Program of China under Grant No. 2016YFF0200403, and Young Scientists Fund of National Natural Science Foundation of China under Grant No. 11504248.
\end{acknowledgments}

\appendix

\section{SVD for sub-matrices of the scattering matrix}
\label{app1}

In this appendix, we show that the unity of the scattering matrix leads to the SVD Eq.~\eqref{s2:eq5-2} in the particle-hole basis.

Without loss of generality, we assume the SVD of the sub-matrices $S^{\rm hp}$ and $S^{\rm hp}$ has the form
\begin{eqnarray}
  S^{\rm ph} & = & U^p i \Lambda_{\rm ph} [V^h]^{\dagger}, \nonumber\\
  S^{\rm hp} & = & U^h i \Lambda_{\rm hp} [V^p]^{\dagger}.
  \label{a1:eq1}
\end{eqnarray}
where both $U^{\rm p(h)}$ and $V^{\rm p(h)}$ are unity matrices, while $\Lambda_{\rm ph(hp)}$ is the semi-position-defined diagonal matrix. Note that we have omit the index $\delta$ for clarification.

By using the matrices $U^{\rm p(h)}$ and $V^{\rm p(h)}$, we can perform a unity transform over the scattering matrix as
\begin{eqnarray}
  \tilde{S} & = & \left(\begin{tabular}{cc}
                        $U^{\rm p}$ & $0$\\
                        $0$ & $U^{\rm h}$\\
                  \end{tabular}\right)^{\dagger} 
                  S \left(\begin{tabular}{cc}
                        $V^{\rm p}$ & $0$\\
                        $0$ & $V^{\rm h}$\\
                  \end{tabular}\right) \nonumber\\
            & = & \left(\begin{tabular}{cc}
                        $U^{\rm p}$ & $0$\\
                        $0$ & $U^{\rm h}$\\
                  \end{tabular}\right)^{\dagger} 
             \left(\begin{tabular}{cc}
                         $S^{\rm pp}$ & $S^{\rm ph}$\\
                        $S^{\rm hp}$ & $S^{\rm hh}$\\
                   \end{tabular}\right)
             \left(\begin{tabular}{cc}
                        $V^{\rm p}$ & $0$\\
                        $0$ & $V^{\rm h}$\\
                  \end{tabular}\right) \nonumber\\
             & = & \left(\begin{tabular}{cc}
                        $[U^{\rm p}]^{\dagger} S^{\rm pp} V^{\rm p}$ & $i \Lambda_{\rm ph}$\\
                        $i \Lambda_{\rm hp}$ & $[U^{\rm h}]^{\dagger} S^{\rm hh} V^{\rm h}$\\
                  \end{tabular}\right).						
  \label{a1:eq2}
\end{eqnarray}

By using the unity of the scattering matrix, one has
\begin{widetext}
\begin{align} 
\tilde{S} \tilde{S}^{\dagger} = \left(\begin{tabular}{cc}
                        $\Lambda^2_{\rm ph} + [U^{\rm p}]^{\dagger} S^{\rm pp} [S^{\rm pp}]^{\dagger} U^{\rm p}$ & $i \Lambda_{\rm ph} [V^{\rm h}]^{\dagger} [S^{\rm hh}]^{\dagger} U^{\rm h} - i [U^{\rm p}]^{\dagger} [S^{\rm pp}]^{\dagger} V^{\rm p} \Lambda_{\rm hp}$\\
                        $i \Lambda_{\rm hp} [V^{\rm p}]^{\dagger} [S^{\rm pp}]^{\dagger} U^{\rm p} -i [U^{\rm h}]^{\dagger} S^{\rm hh} V^{\rm h} \Lambda_{\rm ph}$ & $\Lambda^2_{\rm hp} + [U^{\rm h}]^{\dagger} S^{\rm hh} [S^{\rm hh}]^{\dagger} U^{\rm h}$\\
                  \end{tabular}\right) = \left( \begin{tabular}{cc}
                  							$I$ & $0$\\
											$0$ & $I$\\
                  						 \end{tabular} \right),						
\label{a1:eq3}
\end{align}
\begin{align} 
\tilde{S}^{\dagger} \tilde{S}  = \left(\begin{tabular}{cc}
                        $\Lambda^2_{\rm hp} + [V^{\rm p}]^{\dagger} [S^{\rm pp}]^{\dagger} S^{\rm pp} V^{\rm p}$ & $- i \Lambda_{\rm hp} [U^{\rm h}]^{\dagger} S^{\rm hh} V^{\rm h} + i [V^{\rm p}]^{\dagger} [S^{\rm pp}]^{\dagger} U^{\rm p} \Lambda_{\rm ph}$\\
                        $- i \Lambda_{\rm ph} [U^{\rm p}]^{\dagger} S^{\rm pp} V^{\rm p} + i [V^{\rm h}]^{\dagger} [S^{\rm hh}]^{\dagger} U^{\rm h} \Lambda_{\rm hp}$ & $\Lambda^2_{\rm ph} + [V^{\rm h}]^{\dagger} [S^{\rm hh}]^{\dagger} S^{\rm hh} V^{\rm h}$\\
                  \end{tabular}\right) = \left( \begin{tabular}{cc}
                  							$I$ & $0$\\
											$0$ & $I$\\
                  						 \end{tabular} \right),					
\label{a1:eq4}
\end{align}
\end{widetext}
with $I$ being the unit matrix.

From the above equations, one has the relation(from the diagonal parts)
\begin{eqnarray}
  S^{\rm pp} [S^{\rm pp}]^{\dagger} & = & U^p (I - \Lambda^2_{\rm ph}) [U^p]^{\dagger}, \nonumber\\
  \mbox{} [S^{\rm pp}]^{\dagger} S^{\rm pp} & = & V^p (I - \Lambda^2_{\rm hp}) [V^p]^{\dagger},
  \label{a1:eq5}
\end{eqnarray}
suggesting that the SVD of $S^{\rm pp}$ has the form $S^{\rm pp} = U^p \sqrt{I - P^2} [V^p]^{\dagger}$, with $\Lambda_{\rm ph} = \Lambda_{\rm hp} = P$. Similarly, one has also $S^{\rm hh} = U^h \sqrt{I - P^2} [V^h]^{\dagger}$. One can also check that the off-diagonal parts in Eqs.~\eqref{a1:eq3} and~\eqref{a1:eq4} are also satisfied by such SVDs.

\section{Bloch-Messiah reduction}
\label{app2}

In this section, we show how to obtain the many-body wave function of the outgoing state from the the Bloch-Messiah reduction. In this work, we follow the implement of the reduction introduced by Kraus {\em et al.} in Ref.~[\onlinecite{kraus2009}], which show that any pure fermionic Gaussian state can be cast into standard form 
\begin{eqnarray}
| \Psi \rangle = \prod_k ( u_k + v_k c^{\dagger}_{\rm k 1} c^{\dagger}_{\rm k 2}) | \text{vac} \rangle,
\label{a2:eq1}
\end{eqnarray}
with $| \text{vac} \rangle$ being the vacuum state. Such Gaussian state can be totally decided by two block-diagonalized covariance matrices $R$ and $Q$, whose matrix element can be written as
\begin{eqnarray}
  && R_{\rm k\eta, l\eta'} = \frac{i}{2}\langle \Psi | [c_{\rm k \eta}, c^{\dagger}_{\rm l \eta'}] | \Psi \rangle = \delta_{\rm kl} \delta_{\rm \eta \eta'} \frac{i}{2}(1 - 2 |v_k|^2), \nonumber\\
  && Q_{\rm k\eta, l\eta'} = \frac{i}{2}\langle \Psi | [c_{\rm k \eta}, c_{\rm l \eta'}] | \Psi \rangle = \delta_{\rm kl} [\sigma_y]_{\rm \eta \eta'} u^{\ast}_k v_k, 
\label{a2:eq2}
\end{eqnarray}
with $\sigma_y$ representing the Pauli matrix. The above equations indicates that the pure fermionic Gaussian state can be obtained by reducing the covariance matrix into the form of Eq.~\eqref{a2:eq2}.

The above procedure can be incorporated into the scattering theory approach in the particle-hole basis we have introduced in Sec.~\ref{sec2}. To do this, we first note that the covariance matrices $R$ and $Q$ can be related to the pair correlation function $g$ as
\begin{eqnarray}
R & = & i \left(\begin{tabular}{cc}
       		$0$ & $I$\\
			$I$ & $0$\\
    		\end{tabular}\right) \left(\begin{tabular}{cc}
       		$g^{\rm pp}$ & $g^{\rm ph}$\\
			$g^{\rm hp}$ & $g^{\rm hh}$\\
    		\end{tabular}\right) - \frac{i}{2} \left(\begin{tabular}{cc}
       													$I$ & $0$\\
														$0$ & $I$\\
    												\end{tabular}\right), \nonumber\\
Q & = & \frac{i}{2} [ \left(\begin{tabular}{cc}
       		$g^{\rm pp}$ & $g^{\rm ph}$\\
			$g^{\rm hp}$ & $g^{\rm hh}$\\
    		\end{tabular}\right) - \left(\begin{tabular}{cc}
       		$g^{\rm pp}$ & $g^{\rm ph}$\\
			$g^{\rm hp}$ & $g^{\rm hh}$\\
    		\end{tabular}\right)^{T}],												
\label{a2:eq3}
\end{eqnarray}
where the notation $(...)^T$ represents the matrix transpose. Note that here the vacuum state in this case is the direct product of the Fermi sea states of all the leads. For the two-terminal quantum conductor we considered in Sec.~\ref{sec2}, it corresponds to the state $| F_1 \rangle \otimes | F_2 \rangle$. 

Then, we can construct the pair correlation function within the particle-hole basis from the scattering matrix[Eq.~\eqref{s2:eq2}], which has the form
 \begin{eqnarray}
   && g^{\rm pp}_{\rm \lambda n, \lambda' n'} = \langle \Psi_b | b^p_{\rm
                                                           \lambda n} b^p_{\rm
                                                           \lambda' n'} | \Psi_b \rangle = 0, \nonumber\\
   && g^{\rm hh}_{\rm \lambda n, \lambda' n'} = \langle \Psi_b | b^h_{\rm
                                                           \lambda n} b^h_{\rm
                                                           \lambda' n'} | \Psi_b \rangle = 0, \nonumber\\
   && g^{\rm ph}_{\rm \lambda n, \lambda' n'} = \langle \Psi_b | b^p_{\rm
                                                  \lambda n} b^h_{\rm
                                                  \lambda' n'} |
                                                           \Psi_b \rangle \nonumber\\
                                                     &&\hspace{1.65cm} = \sum_{\rm \eta m} S_{\rm \lambda n, \eta m} S^{\ast}_{\rm \lambda' n', \eta m} ( 1 - f_n ), \nonumber\\
   && g^{\rm hp}_{\rm \lambda n, \lambda' n'} = \langle \Psi_b | b^h_{\rm
                                                  \lambda n} b^p_{\rm
                                                  \lambda' n'} | \Psi_b \rangle \nonumber\\
                                                     &&\hspace{1.65cm} = \sum_{\rm \eta m} S^{\ast}_{\rm \lambda n, \eta m} S_{\rm \lambda' n', \eta m}  f_n,
   \label{a2:eq4}
\end{eqnarray}
where the function $f_n$ is the Fermi distribution function in zero
temperature limit, which has the form
\begin{eqnarray}
   f_n & = & \left\{
                   \begin{tabular}{cc}
                     $0$, & $n \ge 0$\\
                     $1$, &$n<0$\\
                   \end{tabular}
   \right..
   \label{a2:eq5}
\end{eqnarray}
Note that we shall suppress the index $\delta$ when no confusion can arise.

By substituting Eq.~\eqref{a2:eq4} into Eq.~\eqref{a2:eq3}, one find that for each $\delta$, the covariance matrix $R$($Q$) can be solely decided via the scattering matrix:
\begin{eqnarray}
R & = & i \left(\begin{tabular}{cc}
       		$r_p$ & $0$\\
			$0$ & $r^T_h$\\
    		\end{tabular}\right) - \frac{i}{2} \left(\begin{tabular}{cc}
       													$I$ & $0$\\
														$0$ & $I$\\
    												\end{tabular}\right), \nonumber\\
Q & = & \frac{i}{2} \left(\begin{tabular}{cc}
       		$0$ & $q$\\
			$-[q]^T$ & $0$\\
    		\end{tabular}\right),	
   \label{a2:eq6}
\end{eqnarray}
where
\begin{eqnarray}
r_p & = & S^{\rm pp} [S^{\rm pp}]^{\dagger}, \nonumber\\
r_h & = & S^{\rm hh} [S^{\rm hh}]^{\dagger}, \nonumber\\
q   & = & S^{\rm pp} [S^{\rm hp}]^{\dagger} - S^{\rm ph} [S^{\rm hh}]^{\dagger}.
\label{a2:eq7}
\end{eqnarray}

Finally, by using the SVD for the sub-matrices of the scattering matrix, we finds that the covariance matrices $R$ and $Q$ can be bring into a block-diagonalized form:
\begin{eqnarray}
&& R = \frac{i}{2} U^{\dagger} \Big[ 2 \left(\begin{tabular}{cc}
       									$P$ & $0$\\
										$0$ & $P$\\
    		\end{tabular}\right) - \left(\begin{tabular}{cc}
       									$I$ & $0$\\
										$0$ & $I$\\
    								\end{tabular}\right) \Big] U, \label{a2:eq8}\\
&& Q = U^T \left(\begin{tabular}{cc}
       									$0$ & $\sqrt{(I-P)P}$\\
										$-\sqrt{(I-P)P}$ & $0$\\
    							\end{tabular}\right) U.\nonumber
\end{eqnarray}
Within such block-diagonalized form, the wave function can be constructed following Eqs.~\eqref{a2:eq1} and~\eqref{a2:eq2} for each $\delta$. Since all the operators for different $\delta$ are decoupled, the many-body state is just the direct product of these wave functions, this leads to the expression Eqs.~\eqref{s2:eq9} and~\eqref{s2:eq10} in Sec.~\ref{sec2}.

\section{Quasi-particle representation}
\label{app3}

To calculate expectation values of various physical quantities from the many-body state, it is convenient to work with the quasi-particle representation give in Eqs.~\eqref{s2:eq11} and~\eqref{s2:eq12}. From Eq.~\eqref{s2:eq11}, one can see that the only nonzero pair correlation function within the quasiparticle operators is 
\begin{eqnarray}
\langle \Psi_b | \gamma_{\rm k \pm}(\delta) \gamma^{\dagger}_{\rm k \pm}(\delta) | \Psi_b \rangle & = & 1.
\label{a3:eq0}
\end{eqnarray}
From Eq.~\eqref{s2:eq12}, one has
\begin{eqnarray}
&&\hspace{-0.85cm} b^p_n(\delta) = - \sum_k U^p_{\rm nk}(\delta) [\sqrt{1-p_k(\delta)} \gamma_{\rm k +}(\delta) + i\sqrt{p_k} \gamma^{\dagger}_{\rm k -}(\delta)], \label{a3:eq1}\\
&&\hspace{-0.75cm} b^h_n(\delta) = - \sum_k [U^h_{\rm nk}(\delta)]^{\dagger} [\sqrt{1-p_k(\delta)} \gamma_{\rm k -}(\delta) - i\sqrt{p_k} \gamma^{\dagger}_{\rm k +}(\delta)].\nonumber
\end{eqnarray}
The above expressions can be used to calculate the expectation value needed by the Wigner function, for example
\begin{eqnarray}
&& \langle \Psi_b | [b^p_n(\delta)]^{\dagger} b^p_{n'}(\delta) | \Psi_b \rangle \nonumber\\
&& = \sum_{\rm kk'} [U^p_{\rm nk}(\delta) ] U^p_{\rm n'k'}(\delta) p_k(\delta) \delta_{\rm kk'}.
\label{a3:eq2}
\end{eqnarray}

\bibliographystyle{apsrev4-1}
\bibliography{refs}{}

\end{document}